\documentclass[aps,pra,showpacs,amssymb,nofootinbib,superscriptaddress,twocolumn]{revtex4-1}
\usepackage[ansinew]{inputenc}
\usepackage{bbm}
\usepackage{bm}
\usepackage{amsbsy}
\usepackage{amsthm}
\usepackage{amssymb}
\usepackage{amsfonts}
\usepackage{amsmath}
\usepackage{dsfont} 
\usepackage{graphicx} 
\usepackage{epsfig}
\usepackage{epstopdf}
\usepackage{dsfont}
\usepackage[bookmarks=true,colorlinks]{hyperref}
\usepackage[depth=-2,view={FitH 400},open,openlevel=2,atend]{bookmark}
\usepackage{color}
\usepackage[figure,table]{hypcap}
\usepackage{enumerate}

\hypersetup{
	bookmarksnumbered,
	pdfstartview={FitH},
	citecolor={darkgreen},
	linkcolor={darkred},
	urlcolor={darkblue},
	pdfpagemode={UseOutlines}
    }
\definecolor{darkgreen}{RGB}{50,190,50}
\definecolor{darkblue}{RGB}{0,0,190}
\definecolor{darkred}{RGB}{238,0,0}
\usepackage{soul}
\newcommand{\pr}{^{\prime}}
\newcommand{\prpr}{^{\prime\prime}}
\newcommand{\ket}[1]{\ensuremath{\left|\right.\!{#1}\!\left.\right\rangle}}

\newcommand{\bra}[1]{\ensuremath{\left\langle\right.\!{#1}\!\left.\right|}}

\newcommand{\scpr}[2]{\ensuremath{\left\langle\right.\hspace*{-1pt} #1 \hspace*{-1pt}\left|\right.\hspace*{-1pt} #2 \hspace*{-1pt}\left.\right\rangle}}

\newcommand{\comm}[2]{\ensuremath{\left[\right.\! #1 \,, #2 \!\left.\right]}}

\newcommand{\bn}[1]{\ensuremath{\boldsymbol{b}_{\hspace*{0.2pt}\protect\raisebox{-1.5pt}{\scriptsize{$ #1 $}}}}}
\newcommand{\bdn}[1]{\ensuremath{\boldsymbol{b}^{\dagger}_{\hspace*{0.2pt}\protect\raisebox{-0.5pt}{\scriptsize{$ #1 $}}}}}
\newcommand{\psistepij}[2]{\ensuremath{\ket{\hspace*{-0.5pt}\psi_{#1}^{\protect\raisebox{-0.5pt}{\tiny{(#2)}}}\hspace*{-0.5pt}}_{\hspace*{-0.8pt}r}}}
\newcommand{\psistepijbra}[2]{\ensuremath{\bra{\hspace*{-0.5pt}\psi_{#1}^{\protect\raisebox{-0.5pt}{\tiny{(#2)}}}\hspace*{-0.5pt}}}}
\newcommand{\psistepijoverlap}[2]{\ensuremath{_{\,r\!\!}\scpr{\hspace*{-0.5pt}0}{\hspace*{-0.5pt}\psi_{#1}^{\protect\raisebox{-0.5pt}{\tiny{(#2)}}}\hspace*{-0.5pt}}_{\hspace*{-0.8pt}r}}}
\newcommand{\Fmnk}[2]{\ensuremath{\mathcal{F}_{#1}^{\protect\raisebox{-0.5pt}{\tiny{(#2)}}}}}

\newcommand{\tr}{\textnormal{Tr}}
\newcommand{\djj}{d\kern-0.4em\char"16\kern-0.1em}

\begin{document}

\vspace*{-4mm}
\hypertarget{title}{}
\vspace{4mm}
\title{Coherent controlization using superconducting qubits}
\bookmark[named=Coherent controlization using superconducting qubits,level=-2,dest=title]{Coherent controlization using superconducting qubits}
\author{Nicolai Friis}
\email{nicolai.friis@uibk.ac.at}
\affiliation{
Institute for Theoretical Physics, University of Innsbruck,
Technikerstra{\ss}e 21a,
A-6020 Innsbruck,
Austria}
\author{Alexey A. Melnikov}
\affiliation{
Institute for Theoretical Physics, University of Innsbruck,
Technikerstra{\ss}e 21a,
A-6020 Innsbruck,
Austria}
\affiliation{
Institute for Quantum Optics and Quantum Information,
Austrian Academy of Sciences,
Technikerstra{\ss}e 21a,
A-6020 Innsbruck,
Austria}
\author{Gerhard Kirchmair}
\affiliation{
Institute for Quantum Optics and Quantum Information,
Austrian Academy of Sciences,
Technikerstra{\ss}e 21a,
A-6020 Innsbruck,
Austria}
\affiliation{
Institute for Experimental Physics,
University of Innsbruck,
Technikerstra{\ss}e 25,
A-6020 Innsbruck,
Austria
}
\author{Hans J. Briegel}
\affiliation{
Institute for Theoretical Physics, University of Innsbruck,
Technikerstra{\ss}e 21a,
A-6020 Innsbruck,
Austria}

\date{\today}

\begin{abstract}
\vspace*{-5mm}
\hypertarget{sec:abstract}{}
\vspace{4mm}
\bookmark[named=Abstract,level=-1,dest=sec:abstract]{Abstract}
Coherent controlization, i.e., coherent conditioning of arbitrary single- or multi-qubit operations on the state of one or more control qubits, is an important ingredient for the flexible implementation of many algorithms in quantum computation. This is of particular significance when certain subroutines are changing over time or when they are frequently modified, such as in decision-making algorithms for learning agents. We propose a scheme to realize coherent controlization for any number of superconducting qubits coupled to a microwave resonator. For two and three qubits, we present an explicit construction that is of high relevance for quantum learning agents. We demonstrate the feasibility of our proposal, taking into account loss, dephasing, and the cavity self-Kerr effect.
\end{abstract}

\maketitle
\vspace{-12mm}
\hypertarget{sec:intro2}{}
\vspace{6mm}
\section{Introduction}\label{sec:intro}
\bookmark[named=I. Introduction,level=-1,dest=sec:intro2]{I. Introduction}

The ability to coherently control and manipulate individual quantum systems lies at the heart of modern quantum technologies and applications in quantum information~\cite{Haroche2013,Wineland2013,NielsenChuang2000}. Any quantum computation can be realized as~a sequence of elementary quantum gates~\cite{BarencoBennettCleveDiVincenzoMargolusShorSleatorSmolinWeinfurter1995}, which are highly-controlled quantum interactions of few qubits at~a time, and quantum measurements. Prominent applications include, e.g., quantum algorithms for efficient factoring~\cite{DeutschJozsa1992,Shor1994} and quantum simulation~\cite{Feynman1982,BlochDalibardNascibene2012,BlattRoos2012,AspuruGuzikWalther2012,HouckTuereciKoch2012}. More recently, applications of quantum algorithms to certain problems in machine learning, including data classification~\cite{RebentrostMohseniLloyd2014} and search engine ranking~\cite{GarneroneZanardiLidar2012,PaparoMartinDelgado2012}, have been proposed. Other recent proposals, which are of particular interest for the current paper, are the quantum-enhanced deliberation of learning agents in the context of quantum artificial intelligence~\cite{PaparoDunjkoMakmalMartinDelgadoBriegel2014,BriegelDeLasCuevas2012} and the notion of autonomous and adaptive devices for quantum information processing~\cite{TierschGanahlBriegel2015}. In parallel to these theoretical developments, the design of experimental implementations of quantum computational architectures in systems such as, e.g., trapped ions~\cite{KielpinskiMonroeWineland2002,HaeffnerRoosBlatt2008,Schindler-Blatt2013,PiltzSriarunothaiVaronWunderlich2014} and optical setups~\cite{KnillLaflammeMilburn2001,KokMunroNemotoRalphDowlingMilburn2007,MartinLopezLaingLawsonAlvarezZhouOBrien2012} has been greatly advanced.

In addition, the progress in controlling superconducting (SC) quantum systems~\cite{DewesOngSchmittLauroBoulantBertetVionEsteve2012,VanLooFedorovLalumiereSandersBlaisWallraff2013,
RisteDukalskiWatsonDeLangeTiggelmanBlanterLehnertSchoutenDiCarlo2013,RochSchwartzMotzoiMacklinVijayEddinsKorotkovWhaleySarovarSiddiqi2014} has significantly strengthened the role of SC qubits (see, e.g.,~\cite{DevoretWallraffMartinis2004,BlaisHuangWallraffGirvinSchoelkopf2004}) as contenders for the realization of quantum computational devices. In particular, 1-D~\cite{Koch-Schoelkopf2007} and 3-D~\cite{Paik-Schoelkopf2011,Rigetti-Steffen2012} transmons \textemdash\ SC qubits impervious to charge noise \textemdash\ appear as promising candidates. Initial studies of complex algorithms and gate operations~\cite{PlantenbergDeGrootHarmansMooij2007,SillanpaepaeParkSimmonds2007}, fault-tolerance~\cite{ReedDiCarloNiggSunFrunzioGirvinSchoelkopf2012,NiggGirvin2013,Barends-Martinis2014}, and hardware efficiency~\cite{LeghtasKirchmairVlastakisSchoelkopfDevoretMirrahimi2013} in such systems further raise the hopes for~a scalable quantum computational architecture using SC circuitry. Moreover, coupling SC qubits via microwave resonators permits access to the domain of cavity QED~\cite{AstafievZagoskinAbdumalikovPashkinYamamotoInomataNakamuraTsai2010,EichlerBozyigitLangSteffenFinkWallraff2011,AstafievInomataNiskanenYamamotoPashkinNakamuraTsai2007}. There, the exceeding level of control over the combined system can be utilized, e.g., to resolve photon number states~\cite{Schuster-Schoelkopf2007}, or to deterministically encode quantum information in the resonator states~\cite{LeghtasEtAl2013,VlastakisEtAl2013}. We shall draw from this rich quantum optics tool-box in the following.

Notwithstanding these developments, a paradigm with respect to which all of the above implementations are typically deemed successful is the realization of the unitaries of a universal set of quantum gates, see, e.g., Refs.~\cite{NielsenChuang2000,DawsonNielsen2006}. While a universal set of gates in principle allows any unitary to be efficiently approximated, there are some tasks for which this approach lacks a certain \emph{flexibility}. For instance, when specific subroutines of an algorithm require modifications in between individual runs. Prominent examples include the period-finding subroutine in Shor's algorithm~\cite{Shor1994}, which is typically used several times for different functions throughout the algorithm, or when the phases corresponding to different unitaries are to be estimated using Kitaev's scheme~\cite{Kitaev1996}. This issue is also of particular importance when quantum subroutines are used in the decision-making of learning agents~\cite{DunjkoFriisBriegel2015}, which update their subroutines based on experience gathered throughout the learning process. In all these cases, a set of unitaries, which are applied conditionally on the states of an ensemble of auxiliary qubits, is modified in subsequent applications of the subroutines. It is hence of significant interest to establish a method of \emph{coherent controlization} \textemdash\ a mapping from a set of unitaries on a target Hilbert space to a single controlled operation on a larger (control \& target) space \textemdash\ that is independent of the chosen set of unitaries.

Unfortunately, it is not possible to design generic quantum circuits that achieve this conditioning independently of the selected unitaries when only single uses of the unitaries in question are permitted~\cite{ThompsonGuModiVedral2013,AraujoFeixCostaBrukner2014,FriisDunjkoDuerBriegel2014}. On the other hand, physical implementations of the unitaries that are to be controlled are typically already conditioned on fixing some degrees of freedom such as spatial locations (a laser beam illuminating an ion; an optical element being placed in the path of a light beam), or resonance frequencies. This practically allows the realization of purpose-built schemes that ``add control" to unspecified unitaries, e.g., in optical setups~\cite{ZhouRalphKalasuwanZhangPeruzzoLanyonOBrien2011}, or trapped ions~\cite{FriisDunjkoDuerBriegel2014,DunjkoFriisBriegel2015}.

Here, we propose a \emph{modular} and \emph{adaptive} implementation of coherent controlization in a superconducting system of transmon~\cite{Koch-Schoelkopf2007} qubits coupled to a microwave resonator. In the dispersive limit, the coupling between these systems can be understood as well-resolved shifts of the cavity frequencies, dependent on the qubit states, or vice versa, shifts of the qubit frequencies conditioned on the cavity state. Based on this principle, our protocol is assembled from unconditional displacements of the cavity mode and qubit operations conditioned on the vacuum state of the resonator, similar as in Refs.~\cite{LeghtasEtAl2013,VlastakisEtAl2013}. We present a detailed construction of our protocol for two and three qubits, and we give a recipe for up-scaling our scheme to an arbitrary number of qubits. As a cornerstone of our investigation, we include an in-depth analysis of effects detrimental to the success of our protocol. For the strongest source of errors, the cavity-self Kerr, we provide analytical estimates of the disturbance, and discuss methods to reduce it. Using numerical simulations to further take into account decoherence effects such as amplitude- and phase-damping, we hence show that our scheme for coherent controlization can be implemented using current superconducting technology.

This paper is organized as follows. In Sec.~\ref{sec:framework} we briefly review the basic concepts for our proposal: A definition for coherent controlization is given in Sec.~\ref{sec:Coherent controlling unknown unitaries}, illustrated by an application to learning agents in quantum artificial intelligence in Sec.~\ref{sec:Coherent Controlization in the Context of Learning Agents}, before we give a short description of the superconducting transmon qubits that we consider here in Sec.~\ref{eq:Superconducting qubits  coupled to microwave resonators}. The conceptual centrepiece, the coherent controlization protocol for two superconducting qubits, is introduced in Sec.~\ref{sec:coherent controlization of two qubits}, where we first discuss the idealized basic protocol, before turning our attention to the influence of the Kerr effect and decoherence. Finally, we extend our protocol to three qubits and beyond in Sec.~\ref{sec:scaling}.

\hypertarget{sec:framework2}{}
\section{Framework}\label{sec:framework}
\bookmark[named=sec:framework,level=-1,dest=sec:framework2]{II. Framework}

\hypertarget{sec:Coherently controlling unknown unitaries2}{}
\subsection{Coherently controlling unknown unitaries}\label{sec:Coherent controlling unknown unitaries}
\bookmark[named=sec:Coherently controlling unknown unitaries,page=2,level=0,dest=sec:Coherently controlling unknown unitaries2]{II. A. Coherently controlling unknown unitaries}

At the heart of many quantum computational algorithms lie subroutines in which operations of choice on a finite-size register of qubits are performed conditionally on the state of~a control qubit. This is the case, for instance, in Kitaev's phase estimation subroutine~\cite{Kitaev1996}, where, the $2^{n-1}$-fold application~$U(\varphi)^{2^{n-1}}$ of~a phase rotation~$U(\varphi)$ is executed only if the $n$-th ancilla qubit is in the state $\ket{1}_{n}$. Similar subroutines feature also in Shor's factoring algorithm~\cite{Shor1994}. When $U(\varphi)$ is specified, the corresponding subroutine can be efficiently approximated by combining operations from a set of universal quantum gates (see, e.g.,~\cite{DawsonNielsen2006}). The decomposition into the universal gates, and their assembly to form the subroutine clearly requires some (classical) computational effort along with (some) knowledge of $U(\varphi)$. This becomes a practical impediment when a device implementing said subroutine is to be used consecutively for different choices of $U(\varphi)$, possibly diminishing any computational speed-up with respect to purely classical devices. In particular, this issue is of crucial interest for the design of quantum-enhanced autonomous learning agents~\cite{PaparoDunjkoMakmalMartinDelgadoBriegel2014}, where the quantum speed-up concerns the deliberation-time, and the agents need to update their subroutines throughout the learning process.

It would hence be highly desirable to have access to fixed global operations, let us call them $A$ and $B$, on the target and control registers which allow turning an unspecified local operation $U(\varphi)$ into its controlled version $\operatorname{ctrl}-U(\phi)$, such that $AU(\varphi)B=\operatorname{ctrl}-U(\phi)$. However, this requirement cannot be met by any fixed~$A$, $B$ for all $U(\varphi)$, see~\cite{AraujoFeixCostaBrukner2014}, and therefore, in particular, generic system independent controlization is not possible when the action of $U(\varphi)$ on the target Hilbert space is unknown. Fortunately, most practical realizations of (unitary) operations are not strictly local with respect to the target Hilbert space, but are already conditioned on some additional degrees of freedom. For instance, for quantum information encoded in photons, the optical elements must be placed in the beam path, conditioning the operations on spatial degrees of freedom. Another example are laser pulses driving transitions between qubit states encoded in ions, which must be at resonance, conditioning the transformations on the correct frequency. Such implicit conditioning on additional degrees of freedom can be exploited to ``add control" to unitaries that are unknown in the sense specified above~\cite{ZhouRalphKalasuwanZhangPeruzzoLanyonOBrien2011,FriisDunjkoDuerBriegel2014}.

Following Ref.~\cite{DunjkoFriisBriegel2015}, we shall refer to mappings from a set of operations $\left\{U_{i}\right\}$ on the target Hilbert space $\mathcal{H}_{t}$ to operations~$U$ on the joint Hilbert space $\mathcal{H}_{c}\otimes\mathcal{H}_{t}$ of the control and target systems, such that
\begin{align}
    U\ket{i}_{c}\ket{\psi}_{t}  &=\,\sum\limits_{i}\ket{i}_{c}U_{i}\ket{\psi}_{t}\ \ \forall \ket{\psi}_{t}\in\mathcal{H}_{t}
\end{align}
for some (orthonormal) basis $\left\{\ket{i}\right\}_{c}$ of $\mathcal{H}_{c}$, as well as to any specific physical realization of such mappings as \emph{coherent controlization}. In the following, we shall discuss how general coherent controlization can be implemented in superconducting qubits, providing a modular and adaptive architecture for quantum computational tasks. For~a detailed analysis of the \emph{feasibility} and \emph{scalability} of our proposal, we provide~a concrete example for an application of coherent controlization in the decision process of learning agents, where the adaptive character of our proposal is of particular relevance.

\hypertarget{sec:Coherent Controlization in the Context of Learning Agents2}{}
\subsection{Coherent controlization in the context of learning agents}\label{sec:Coherent Controlization in the Context of Learning Agents}
\bookmark[named=sec:Coherent Controlization in the Context of Learning Agents,level=0,dest=sec:Coherent Controlization in the Context of Learning Agents2]{II. B. Coherent controlization in the context of learning agents}

In the model of projective simulation (PS)~\cite{BriegelDeLasCuevas2012}, an autonomous learning agent draws upon previous experience to simulate its future situation in a given (and partially unknown) environment. The centrepiece of such an agent, which is also equipped with sensors (to receive perceptual input, \emph{percepts}, from the environment) and actuators (enabling it to act on and change the environment) is a specific type of memory (ECM)~\cite{BriegelDeLasCuevas2012}. In abstract terms, the memory is represented by a space of clips that can represent percepts, actions, and combinations thereof. After receiving sensory input, the PS agent initiates a random walk within the clip space to find an action. Under a given reward scheme, the agent's choices have consequences that modify and update its memory, it \emph{learns}. Throughout the learning process, the agent must hence be able to adjust its deliberation according to its experience, which entails updating the stochastic matrix $P=(p_{ij})$ of transition probabilities of the random walk in the space of memory clips. Recently, the PS model has been shown to perform competitively in typical artificial intelligence benchmark tasks~\cite{MelnikovMakmalBriegel2014}, and, further,
that its memory structure provides a dynamic framework for generalization~\cite{MelnikovMakmalDunjkoBriegel2015}.

A particular variant of the PS that we shall focus on now, is reflecting projective simulation (RPS)~\cite{PaparoDunjkoMakmalMartinDelgadoBriegel2014}. In the RPS framework, the random walk in the memory space is continued until the underlying Markov chain $P=(p_{ij})$ is (nearly) mixed, and actions are then \mbox{sampled} according to the stationary distribution $\pi$, where $P\pi=\pi$. The quantized version of the RPS yields a quadratic speed-up with respect to its classical counterpart, both in the number of calls to~$P$ needed to mix the chain, and in the number of samples until an action is obtained. The Szegedy-type quantum random walk involved in this procedure requires several levels of coherent controlization. At the lowest level of this nested scheme of adding control, one encounters a set of unitaries $\{U_{j}\}_{j=1}^{n}$ that encode the $n\times n$ stochastic matrix~$P$ of an $n$-clip network. That is, the first column of $U_{j}$ has real and positive entries $\sqrt{p_{ij}}$ $(i=1,2,\ldots,n)$, and each such \emph{probability unitary} $U_{j}$ may thus be parameterized by $(n-1)$ real angles~$\theta_{j,1},\ldots,\theta_{j,n-1}$. After each step of the learning process, the matrix~$P$ is updated according to the rewards that may have been incurred, requiring also updates of the $U_{j}$. In the spirit of adaptiveness of the agent's design it is desirable that these updates can be carried out by directly updating the angles $\theta_{j,k}$ in an otherwise fixed hardware. A method~\cite{DunjkoFriisBriegel2015} for realizing this requirement is~a nested construction of coherent controlization for $\log_{2}(n)$ qubits, where control is added to $(n-1)$ single-qubit $Y$-rotations $U(\theta_{j,k})$ for each $U_{j}$.

For instance, for two-qubits, the unitary $U(\theta_{1})$ is unconditionally applied on the control qubit, followed by the applications of $U(\theta_{2})$ and $U(\theta_{3})$ conditioned on the control qubit being in the states $\ket{1}_{c}$ or $\ket{0}_{c}$, respectively, see Fig.~\ref{fig:coherent controlization circuit diagram}~(b). For three qubits, a pair of two-qubit subroutines of the form just described replaces the two conditional single-qubit operations, see Fig.~\ref{fig:coherent controlization three qubits}~(a), and so forth. This construction already entails the conditioning of single-qubit operations on all subspaces of the control qubits. We shall therefore consider the implementation of the probability unitaries $U_{j}$ in superconducting qubits as~a representative example that demonstrates the feasibility of coherent controlization. In order to proceed, we shall next give a brief overview of the properties of the superconducting qubit systems suitable for our purposes.

\hypertarget{sec:Superconducting qubits coupled to microwave resonators2}{}
\subsection{Superconducting qubits coupled to microwave resonators}\label{eq:Superconducting qubits  coupled to microwave resonators}
\bookmark[named=sec:Superconducting qubits coupled to microwave resonators,level=0,dest=sec:Superconducting qubits coupled to microwave resonators2]{II. C. Superconducting qubits coupled to microwave resonators}

The physical system that we consider in our proposal is an array of superconducting transmon qubits~\cite{Koch-Schoelkopf2007} coupled to a microwave resonator. For the reader unfamiliar with the principal design of such a system, let us give an intuitive example. Consider a~basic superconducting $LC$\textendash circuit, which may be thought of as the realization of~a quantum mechanical harmonic oscillator, where charge and flux take the role of the canonically conjugate variables. Via the non-linearity of Josephson-junctions, an anharmonicity can be introduced into the system, which modifies the otherwise equal energy level spacing. This allows one to frequency-address transitions between two chosen levels (typically the two lowest-lying levels), thus forming~a qubit. For the practical realization of such a macroscopic qubit, several options, such as charge and flux qubits, are available and we direct the reader to pedagogic reviews (see, e.g., Ref.~\cite{DevoretWallraffMartinis2004}) for more details on their differences.

Here, we shall focus on the \emph{transmon qubit}, introduced in Ref.~\cite{Koch-Schoelkopf2007}, which is in its design similar to usual charge qubits, in the sense that two superconducting islands are connected via~a Josephson junction with associated Josephson energy $E_{J}$. In some setups two Josephson junctions can be used to form~a dc-SQUID. This leaves the possibility to modify the Josephson energy $E_{J}$ of the junctions by threading the SQUID with an external magnetic flux. Besides $E_{J}$, the energy levels of such charge qubits are determined by the charging energy $E_{C}$ of the superconducting island. The departure of the transmon from other designs lies in the introduction of~a large shunting capacitance parallel to the dc-SQUID, which drastically reduces $E_{C}$. The transmon qubit is hence operated in~a regime where $E_{J}\gg E_{C}$, which leads to an exponential decrease of the charge dispersion in $E_{J}/E_{C}$, while the anharmonicity is only diminished polynomially in this ratio. In other words, the qubit levels remain addressable by frequency selection, while their sensitivity to environmentally induced charge noise is practically removed, which significantly improves qubit coherence times ($\gtrsim 100~\mu$s)~\cite{Paik-Schoelkopf2011,Rigetti-Steffen2012}.

For our investigation, several such transmon qubits shall be considered to be capacitively coupled to~a superconducting resonator. To good approximation (see, e.g., the \hyperref[sec:appendix]{Appendix}), the Hamiltonian for this dispersive interaction may be written as~\cite{Nigg-Girvin2012,BourassaBeaudoinGambettaBlais2012}
\begin{align}
    H/\hbar   &=\,\omega_{r}\boldsymbol{a}^{\dagger}\boldsymbol{a}\,+\,\sum\limits_{i}\omega_{q_{i}}\bdn{i}\bn{i}\,
    -\,\sum\limits_{i}\chi_{q_{i}r}\,\boldsymbol{a}^{\dagger}\boldsymbol{a}\,\bdn{i}\bn{i}
    \,\nonumber\\
    &\ \ -\,\sum\limits_{i}\frac{\chi_{q_{i}q_{i}}}{2}\bigl(\bdn{i}\bn{i}\bigr)^{\!2}\,
    -\,\frac{\chi_{rr}}{2}\bigl(\boldsymbol{a}^{\dagger}\boldsymbol{a}\bigr)^{\!2}\,,
    \label{eq:supercond qubits Hamiltonian}
\end{align}
where $\omega_{r}$ and $\omega_{q_{i}}$ are the (angular) frequencies of the resonator mode and the $i$-th qubit, respectively, with the corresponding ladder operators $\boldsymbol{a},\boldsymbol{a}^{\dagger}$, $\bn{i},\bdn{i}$. For the remainder of this paper, higher excitation numbers are ignored for the qubit-modes. Due to the coupling of the qubits to the resonator, the latter also acquires an anharmonicity, which causes the undesired self-Kerr effect~\cite{KirchmairEtAl2013} represented by $\chi_{rr}$. The constants $\chi_{rr}$, $\chi_{q_{i}q_{i}}$, and the cross-Kerr coefficient $\chi_{q_{i}r}$ are hence not independent. For instance, when only one qubit is present, $q_{i}=q$, the relation between these parameters is $\chi_{rr}=\chi_{qr}^{2}/(4\chi_{qq})$ in the dispersive limit~\cite{Nigg-Girvin2012}. Typical values for these parameters that we will consider in the two-qubit case are $\chi_{q_{i}r}/(2\pi)\approx1-5$~MHz, and $\chi_{qq}/(2\pi)\approx300$~MHz (where we have assumed $\chi_{q_{1}q_{1}}=\chi_{q_{2}q_{2}}=\chi_{qq}$), corresponding to cavity anharmonicities $\chi_{rr}/(2\pi)$ roughly between $0.7-17$~kHz.

The terms proportional to $\chi_{q_{i}r}$ can be interpreted as conditional frequency shifts: Depending on the number of excited qubits, the resonator frequency $\omega_{r}$ can be regarded as shifted by the sum of the corresponding values $\chi_{q_{i}r}$. Conversely, the frequency of the $i$-th qubit can be considered as being shifted by the product of $\chi_{q_{i}r}$ and the number of photons in the cavity. In the strong dispersive regime~\cite{VlastakisEtAl2013}, the spectral lines of the cavity for different qubit states are well-resolved, and, \textit{vice versa}, so are the qubit transition frequencies for different numbers of excitations of the resonator. Both of these points of view will play an important role in our protocol for coherent controlization that we will introduce next.

\hypertarget{sec:Coherent controlization for two superconducting qubits2}{}
\section{Coherent controlization for two superconducting qubits}\label{sec:coherent controlization of two qubits}
\bookmark[named=sec:Coherent controlization for two superconducting qubits,level=-1,dest=sec:Coherent controlization for two superconducting qubits2]{III. Coherent controlization for two superconducting qubits}

In the protocol for coherent controlization that we propose here, the phase space of the resonator mode serves as~a bus between the qubits. It enables the conditioning of operations on particular subspaces of the qubit Hilbert space by separating the resonator states corresponding to different qubit subspaces in phase space. The mechanism for this separation is the free time evolution of coherent states with different frequencies. Recall that the resonance frequency of the cavity depends on the state of the qubits. Similar to the procedures used in~\cite{LeghtasEtAl2013,VlastakisEtAl2013,CavesShaji2010}, the following operations are employed for our protocol:
\begin{itemize}
    \item \textbf{Unconditional displacements} $D_{\alpha}$: A very short pulse (a few ns), that is, sufficiently broad in frequency (of width $\gg\sum_{i}\chi_{q_{i}r}$), so as not to distinguish between the state-dependent frequencies of the resonator, displaces the cavity state independently of the state of the qubits.
    \item \textbf{Free time evolution} $U(t)$: During the free time evolution of the cavity mode, governed by the corresponding parts of the Hamiltonian in~(\ref{eq:supercond qubits Hamiltonian}), coherent states of the resonator corresponding to different qubit states rotate in phase space at different speeds. Appropriate waiting periods can hence be used to separate or recombine different coherent states.
    \item \textbf{Conditional qubit operations:} When the (average) photon number in the resonator is $\bar{n}$, the (mean) qubit frequencies $\omega_{q_{i}}$ are shifted to $\omega_{q_{i}}-\chi_{q_{i}r}\bar{n}$ with~a spread $\chi_{q_{i}r}\sqrt{\bar{n}}$. Addressing the qubits with signals sufficiently narrow in frequency around $\omega_{q_{i}}$ therefore conditions the single-qubit operations on the cavity vacuum state $\ket{0}_{r}$. Unconditional qubit operations can be realized by appropriately broad (or multi-frequency) pulses.
\end{itemize}
With these operations available we shall now specialize to the case of two qubits.

\hypertarget{sec:two qubit protocol2}{}
\subsection{Ideal two-qubit protocol}\label{sec:two qubit protocol}
\bookmark[named=sec:two qubit protocol,level=0,dest=sec:two qubit protocol2]{III. A. Ideal two-qubit protocol}

\begin{figure*}[ht!]
    \setlength{\fboxsep}{5pt}\setlength{\fboxrule}{1pt}\framebox{
        \parbox{0.98\textwidth}{
    \begin{minipage}{0.47\textwidth}
        \textbf{(a)}\ \ \textbf{Ideal two-qubit protocol}
        \hspace*{-1mm}
        \begin{enumerate}[(i)]
            \item \label{step i}
                The first (unconditional) single-qubit unitary $U(\theta_{1})$ is applied to qubit~$2$, mapping the initially (pure) state $\ket{\psi}_{q}\ket{0}_{r}$ to $\bigl(c_{00}\ket{00}_{q}+c_{01}\ket{01}_{q}+c_{10}\ket{10}_{q}+c_{11}\ket{11}_{q}\bigr)\ket{0}_{r}$,  where the coefficients $c_{mn}=c_{mn}(\theta_{1})$ depend on the initial state and $U(\theta_{1})$.
            \item \label{step ii} An unconditional displacement $D_{\alpha}$ maps $\ket{0}_{r}$ to $\ket{\alpha}_{r}$.
            \item \label{step iii} In the frame rotating with $\omega_{00}=\omega_{r}$, the coherent state components for the qubit states rotate with (angular) frequencies $0,\Delta\omega,2\Delta\omega$, and $3\Delta\omega$ for $\ket{00}_{q}$, $\ket{01}_{q}$, $\ket{10}_{q}$, and $\ket{11}_{q}$, respectively. That is, the time evolution $U_{I}=e^{-iH_{\hspace*{-0.5pt}I}t/\hbar}$ is governed by the cross-Kerr interaction term $H_{\hspace*{-0.5pt}I}$. A waiting period of $\Delta t=\pi/\Delta\omega$ hence results in the state $\bigl(c_{00}\ket{00}_{q}+c_{10}\ket{10}_{q}\bigr)\ket{\alpha}_{r}+\bigl(c_{01}\ket{01}_{q}+c_{11}\ket{11}_{q}\bigr)\ket{-\alpha}_{r}$, separating the subspaces for $\ket{0}_{q_{2}}$ and $\ket{1}_{q_{2}}$.
            \item \label{step iv} Another unconditional displacement $D_{\alpha}$ shifts the coherent state components for the subspace $\ket{1}_{q_{2}}$ to the vacuum, $\ket{-\alpha}_{r}\mapsto\ket{0}_{r}$, while $\ket{\alpha}_{r}\mapsto\ket{2\alpha}_{r}$.
        \end{enumerate}
    \end{minipage}
    \hspace*{3mm}
    \begin{minipage}{0.47\textwidth}
        \begin{enumerate}[(i)]\addtocounter{enumi}{4}
            \item \label{step v} During another waiting period of duration $\Delta t$ the second single-qubit unitary $U(\theta_{2})$ is applied conditionally on the cavity being in the vacuum. Meanwhile, the coherent state components $\ket{00}_{q}\ket{\alpha}_{r}$ and $\ket{10}_{q}\ket{\alpha}_{r}$ 
                complete $0$ and $1$ revolution, respectively, returning to the initial position of step~(\ref{step v}).
            \item \label{step vi} An unconditional displacement by $-2\alpha$ exchanges the roles of the subspaces $\ket{0}_{q_{2}}$ and $\ket{1}_{q_{2}}$.
            \item \label{step vii} During the penultimate waiting period of $\Delta t$, which rotates $\bigl(\ket{01}_{q}+\ket{11}_{q}\bigr)\ket{-2\alpha}_{r}$ to
                $\bigl(\ket{01}_{q}+\ket{11}_{q}\bigr)\ket{2\alpha}_{r}$, the third single-qubit operation $U(\theta_{3})$ is applied conditionally on the cavity vacuum state.
            \item \label{step viii} An unconditional displacement $D_{-\alpha}$ shifts $\ket{2\alpha}_{r}$ to $\ket{\alpha}_{r}$, and $\ket{0}_{r}$ to $\ket{-\alpha}_{r}$.
            \item \label{step ix} The last period of free time evolution by $\Delta t$ rotates all phase space components to $\ket{-\alpha}_{r}$, disentangling the cavity from the qubits.
            \item \label{step x} The final unconditional displacement $D_{\alpha}$ returns the cavity to the vacuum state.
\end{enumerate}
    \end{minipage}}}
    \begin{minipage}{0.45\textwidth}
    \hspace*{-6mm}\textbf{(b)}\hspace*{10mm}\includegraphics[width=0.7\textwidth]{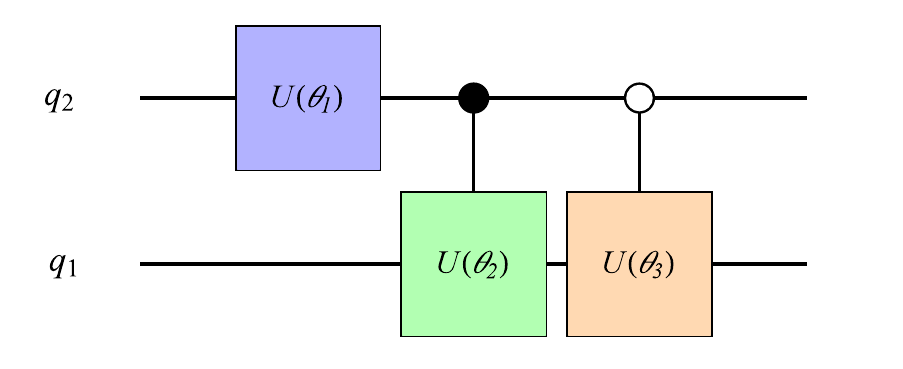}
    \textbf{(c)}\includegraphics[width=0.9\textwidth]{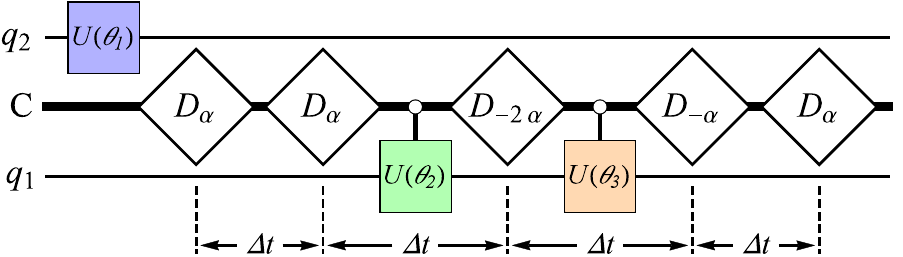}
    \end{minipage}
    \begin{minipage}{0.53\textwidth}
    \textbf{(d)}\includegraphics[width=0.98\textwidth]{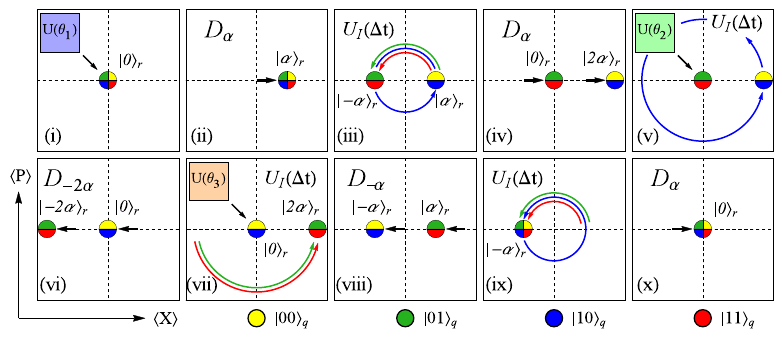}
    \end{minipage}
    \caption{\label{fig:coherent controlization circuit diagram}
    \textbf{Two-qubit protocol for coherent controlization}: The steps (\ref{step i})-(\ref{step x}), listed in~(a) realize the probability unitaries $U_{j}$ (positive, real entries in the first column), represented by the circuit diagram in~(b). The filled dots ``$\bullet$" on the controlled operations indicate that the unitaries on the target are conditioned on the control qubit state~$\ket{1}_{q_{2}}$, while the hollow dots ``$\circ$" denote conditioning on the control qubit state~$\ket{0}_{q_{2}}$, by way of coherent controlization of the three single qubit unitaries $U(\theta_{1})$, $U(\theta_{2})$, and $U(\theta_{3})$. In~(c), an extended circuit diagram is shown, where the upper and lower lines indicate qubits~$q_{2}$ and~$q_{1}$, respectively, while the middle line labelled~C represents the cavity mode. The white rectangles represent unconditional displacements, and the control lines between the cavity and the single-qubit operations indicate operations conditioned on the resonator being in the ground state. The waiting time between the displacements is $\Delta t=\pi/\Delta\omega$. (d) shows the phase space representation of the protocol for the coherent state components corresponding to different qubit states.}
\end{figure*}

Let us first analyze an idealized situation where the cavity Kerr effect due to $H_{\hspace*{-0.5pt}K}/\hbar=-\,\frac{\chi_{rr}}{2}\bigl(\boldsymbol{a}^{\dagger}\boldsymbol{a}\bigr)^{\!2}$ and any loss due to interactions with the environment can be disregarded. Once we have set up our protocol for two qubits in this idealized setting, we shall consider the robustness of the protocol under the influence of the mentioned harmful effects. Irrespective of this restriction, the cross-Kerr interaction term $H_{\hspace*{-0.5pt}I}/\hbar=-\,\sum_{i}\chi_{q_{i}r}\,\boldsymbol{a}^{\dagger}\boldsymbol{a}\,\bdn{i}\bn{i}$ (in the strong dispersive regime) splits the cavity resonance frequency into $4$ well-resolved spectral lines $\omega_{00}$, $\omega_{01}$, $\omega_{10}$, and $\omega_{11}$, corresponding to the two-qubit basis states $\ket{00}_{q}$, $\ket{01}_{q}$, $\ket{10}_{q}$, and $\ket{11}_{q}$. By selecting the qubit-cavity cross-Kerr coefficients $\chi_{q_{1}r}/2=\chi_{q_{2}r}=\Delta\omega\approx1-5\times2\pi$~MHz, the spectral lines are equally spaced, i.e.,
\begin{align}
    \omega_{00} &=\,\omega_{01}+\Delta\omega=\omega_{10}+2\Delta\omega=\omega_{11}+3\Delta\omega=\omega_{r}\,,
\end{align}
which can be abbreviated to $\omega_{mn}=\omega_{r}-(2m+n)\Delta\omega$. With the resonator prepared in the vacuum state $\ket{0}_{r}$, we now construct a protocol that implements coherent controlization for a set of three single-qubit unitaries $\{U(\theta_{i})\}_{i=1,2,3}$ to realize the probability unitary
\begin{small}
\begin{align}
    U   &=\,
    \Bigl(\ket{0}\!\bra{0}_{q_{2}}\otimes U(\theta_{3})_{q_{1}}+\ket{1}\!\bra{1}_{q_{2}}\otimes U(\theta_{2})_{q_{1}}\Bigr)\nonumber\\
    &\ \ \times\bigl(U(\theta_{1})_{q_{2}}\otimes\mathds{1}_{q_{1}}\bigr)\,,
    \label{eq:probability unitary}
\end{align}
\end{small}
represented by the circuit-diagram in Fig.~\ref{fig:coherent controlization circuit diagram}~(b). The protocol, whose steps are detailed in Fig.~\ref{fig:coherent controlization circuit diagram}~(a) and which are illustrated in Fig.~\ref{fig:coherent controlization circuit diagram}~(b) and~(c), can be decomposed entirely into the operations described above.

Given that the initial (unconditional) qubit rotation $U(\theta_{1})$ can be performed very rapidly ($\approx10$~ns), the overall gate time~$T_{2QB}$ is approximately $T_{2QB}\approx4\Delta t=4\pi/\Delta\omega=400$~ns to $2~\mu$s [for $\chi_{q_{2}r}/(2\pi)$ between $5$~MHz and $1$~MHz], compared to typical~\cite{LeghtasEtAl2013} cavity coherence times $\tau_{r}\approx100~\mu$s, and qubit relaxation- and dephasing times of around $20-100~\mu$s. To maintain a high fidelity, especially when the protocol is extended to several qubits (as we shall do in the following), it is desirable to decrease the overall gate time, and hence to increase the qubit-cavity cross-Kerr coefficients $\chi_{q_{i}r}$. However, this can only be done at the expense of an increased cavity self-Kerr term proportional to $\chi_{rr}$, which grows quadratically with the $\chi_{q_{i}r}$. In other words, there is a trade-off between decoherence and imperfections of the gates due to distortions of the coherent-state components (through the Kerr effect). Before we extend our protocol to larger numbers of qubits, we will therefore study the robustness of the two-qubit protocol under these effects.

\hypertarget{sec:Influence of the cavity self-Kerr effect2}{}
\subsection{Influence of the cavity self-Kerr effect}\label{sec:Influence of the cavity self-Kerr effect}
\bookmark[named=sec:Influence of the cavity self-Kerr effect,level=0,dest=sec:Influence of the cavity self-Kerr effect2]{III. B. Influence of the cavity self-Kerr effect}

With increasing coupling between the qubits and the resonator, the influence of the anharmonicities of the transmon qubits on the resonator mode becomes ever stronger. The ensuing cavity Kerr effect distorts (and displaces) the shape of the phase space distributions~\cite{KirchmairEtAl2013}. To estimate the impact on the performance of our protocol, we shall more closely inspect the three stages at which the overlap of the phase space distributions with the target states determines the success of the protocol. First, let us denote the overall state after step $k$ as \begin{align}
    \sum\limits_{m,n=0,1}c_{mn}^{\raisebox{-0.5pt}{\tiny{(k)}}}\psistepij{mn}{k}\ket{mn}_{q}\,.
    \label{eq:step k state}
\end{align}
During steps~(\ref{step iv}) and~(\ref{step vi}), the coherent state components $\psistepij{01}{iv}$, $\psistepij{11}{iv}$ and $\psistepij{00}{vi}$, $\psistepij{10}{vi}$, corresponding to the subspaces $\ket{1}_{q_{2}}$ and $\ket{0}_{q_{2}}$, respectively, are required to significantly overlap with the vacuum to achieve the conditioning of $U(\theta_{2})$ and $U(\theta_{3})$. Finally, after being disentangled from the qubits in step~(\ref{step ix}), all resonator states $\psistepij{mn}{x}$\ $(i,j=0,1)$ should ideally return to the vacuum in step~(\ref{step x}). The resonator states at these various stages can be obtained by applying the displacements and waiting periods as described in the Fig.~\ref{fig:coherent controlization circuit diagram}~(a), where the time evolution is now determined by $H_{\hspace*{-0.5pt}I}+H_{\hspace*{-0.5pt}K}$, with $H_{\hspace*{-0.5pt}K}/\hbar=\,-\frac{\chi_{rr}}{2}\bigl(\boldsymbol{a}^{\dagger}\boldsymbol{a}\bigr)^{\!2}$, i.e.,
\begin{subequations}
\label{eq:resonator states with Kerr}
\begin{align}
    \psistepij{mn}{iv}\ket{\hspace*{-0.8pt}mn\!}_{\!q}    &=D_{\alpha}U_{I\hspace*{-0.5pt},K}(\Delta t)D_{\alpha}\ket{0}_{r}\ket{\hspace*{-0.8pt}mn\!}_{\!q} ,
    \label{eq:resonator states with Kerr step v}\\
    \psistepij{mn}{vi}\ket{\hspace*{-0.8pt}mn\!}_{\!q}    &=D_{-2\alpha}U_{I\hspace*{-0.5pt},K}(\Delta t)\psistepij{mn}{iv}\ket{\hspace*{-0.8pt}mn\!}_{\!q} ,
    \label{eq:resonator states with Kerr step vii}\\
    \psistepij{mn}{x}\ket{\hspace*{-0.8pt}mn\!}_{\!q}     &=D_{\alpha}U_{I\hspace*{-0.5pt},K}(\Delta t)D_{-\alpha}U_{I\hspace*{-0.5pt},K}(\Delta t)\psistepij{mn}{vi}\ket{\hspace*{-0.8pt}mn\!}_{\!q},
    \label{eq:resonator states with Kerr step x}
\end{align}
\end{subequations}
where $U_{I,K}(t)=U_{I}(t)U_{K}(t)=e^{-iH_{\hspace*{-0.5pt}I}t/\hbar} e^{-iH_{\hspace*{-0.5pt}K}t/\hbar}$ and $\Delta t=\pi/\Delta\omega$. Note that the operations $U_{I}$ can be eliminated from~(\ref{eq:resonator states with Kerr}), since $[H_{\hspace*{-0.5pt}I},H_{\hspace*{-0.5pt}K}]=0$ and the displacements $D_{\alpha}=\exp\bigl(\alpha \boldsymbol{a}^{\dagger}-\alpha^{*}\boldsymbol{a}\bigr)$ satisfy $\exp\bigl(i\phi\hspace*{0.5pt} \boldsymbol{a}^{\dagger}\boldsymbol{a}\bigr)D_{\alpha}=D_{e^{i\phi}\alpha}\exp\bigl(i\phi\hspace*{0.5pt} \boldsymbol{a}^{\dagger}\boldsymbol{a}\bigr)$. Acting on~a state with fixed $m,n$, one therefore has
\begin{align}
    U_{I}(\Delta t)D_{\alpha}   &=
    e^{-i\omega\pr_{mn}\Delta t\hspace*{0.5pt}\boldsymbol{a}^{\hspace*{-0.5pt}\dagger}\!\boldsymbol{a}}D_{\alpha}=
    D_{e^{i(2m+n)\pi}\alpha}e^{i(2m+n)\pi\hspace*{0.5pt}\boldsymbol{a}^{\hspace*{-0.5pt}\dagger}\!\boldsymbol{a}},
    \label{eq:commuting displacements and time evolution}
\end{align}
where the rotating frame frequencies are $\omega\pr_{mn}=\omega_{mn}-\omega_{r}=-(2m+n)\Delta\omega$. The $U_{I}(\Delta t)$ can hence be easily commuted with the displacements~$D_{\alpha}$ (at most changing the signs of the displacement parameters for $(m,n)=(0,1),(1,1)$). In the parameter regime that we have chosen, one further has $\chi_{rr}\Delta t/2\approx6.5\times10^{-3}-3\times10^{-2}\ll1$, where the couplings of the two qubits to the cavity contribute to the cavity Kerr term approximately (see the \hyperref[sec:appendix]{Appendix}) as~$\chi_{rr}\approx(\chi_{q_{1}r}^{2}+\chi_{q_{2}r}^{2})/(4\chi_{qq})$. We may therefore expand $U_{K}(\Delta t)$ into a power series for small $\epsilon\equiv\chi_{rr}\Delta t/2$, i.e.,
\begin{align}
    U_{K}(\Delta t) &=e^{i\epsilon\,(\boldsymbol{a}^{\hspace*{-0.5pt}\dagger}\!\boldsymbol{a})^{2}}=\mathds{1}+i\epsilon(\boldsymbol{a}^{\hspace*{-0.5pt}\dagger}\!\boldsymbol{a})^{2}
    -\frac{\epsilon^{2}}{2}(\boldsymbol{a}^{\hspace*{-0.5pt}\dagger}\!\boldsymbol{a})^{4}+\mathcal{O}(\epsilon^{3}),
\end{align}
where $\mathcal{O}(x)$ is a quantity such that $\mathcal{O}(x)/x$ is bounded in the limit $x\rightarrow0$. Since $D_{-\alpha}=D^{\dagger}_{\alpha}=D^{-1}_{\alpha}$, we can then use the simple identities $D_{-\alpha}\boldsymbol{a}^{\dagger}D_{\alpha}=\boldsymbol{a}^{\dagger}+\alpha^{*}$ and $D_{-\alpha}\boldsymbol{a}D_{\alpha}=\boldsymbol{a}+\alpha$. With some tedious but straightforward algebra we arrive at the estimates for the desired overlap after step~(\ref{step iv}), i.e.,
\begin{align}
    \Fmnk{01}{iv}&=\,\Fmnk{11}{iv}\,=\,1\,-\,\epsilon^{2}\bigl(4\bar{n}^{3}\,+\,6\bar{n}^{2}\,+\,\bar{n}\bigr)\,+\,\mathcal{O}(\epsilon^{3})\,,
    \label{eq:overlap with Kerr step iv}
\end{align}
where $\bar{n}=|\alpha|^{2}$ and we have used the shorthand
\begin{align}
    \Fmnk{mn}{k}    &=|_{q\!\!}\bra{\hspace*{-0.8pt}mn\!}\psistepijoverlap{mn}{k}\ket{\hspace*{-0.8pt}mn\!}_{\!q} |^{2}\,.
    \label{eq:overlap fidelities}
\end{align}
Similarly, we find the overlaps after steps~(\ref{step vi}),
\begin{align}
    \Fmnk{00}{vi}&=\,\Fmnk{10}{vi}\,=\,1\,-\,\epsilon^{2}\bigl(409\bar{n}^{3}\,+\,158\bar{n}^{2}\,+\,9\bar{n}\bigr)+\mathcal{O}(\epsilon^{3}),
    \label{eq:overlap with Kerr step vi}
\end{align}
and for step~(\ref{step x}) for all $m,n=0,1$:
\begin{align}
    \Fmnk{mn}{x}&=1\,-\,\epsilon^{2}\bigl(256\bar{n}^{3}\,+\,136\bar{n}^{2}\,+\,4\bar{n}\bigr)\,+\,\mathcal{O}(\epsilon^{3})\,.
    \label{eq:overlap with Kerr step x}
\end{align}
The explanation for the revival of the fidelity from step~(\ref{step vi}) to~(\ref{step x}) lies in the leading order Kerr effect. As discussed in~\cite[App.~B]{LeghtasEtAl2013}, to linear order in $\epsilon$, the Kerr effect does not distort the shape of the coherent states, but adds an amplitude-dependent rotation and global phase factor to each coherent state, that is
\begin{align}
    e^{i\epsilon(\boldsymbol{a}^{\hspace*{-0.5pt}\dagger}\!\boldsymbol{a})^{2}}\ket{\alpha}_{r} &=\,e^{-i\epsilon\bar{n}^{2}}\,\ket{e^{i\epsilon(2\bar{n}+1)}\alpha}\,+\,\mathcal{O}(\epsilon^{2})\,.
    \label{eq:linear Kerr effect}
\end{align}
Although the phases $e^{-i\epsilon\bar{n}^{2}}$ appear as relative phases between different qubit states, our protocol is designed in such a way that all qubit states acquire the same (and hence a global) phase: in steps~(\ref{step iii}) and~(\ref{step ix}) all components have the same amplitude $|\alpha|$, and the phases picked up by $\ket{00}_{q}$ and $\ket{10}_{q}$ in step~(\ref{step v}) are compensated by those acquired by $\ket{01}_{q}$ and $\ket{11}_{q}$ in step~(\ref{step vii}).

The additional rotation(s) from $\alpha$ to $e^{i\epsilon(2\bar{n}+1)}\alpha$, on the other hand, are only partially compensated in the protocol of Fig.~\ref{fig:coherent controlization circuit diagram}. The coherent state components $\ket{00}_{q}$ and $\ket{10}_{q}$ overshoot their target in steps~(\ref{step iii}) and~(\ref{step v}), i.e., the coherent states are rotated counter-clockwise with respect to their target on the horizontal axis. This means, the displacement $D_{-\alpha}$ in step~(\ref{step viii}) leaves them lagging behind, that is, rotated clockwise w.r.t. their target. This lag is partially compensated by the over-rotation in step~(\ref{step ix}) in the sense that the mismatch in step~(\ref{step iii}) is fully corrected, but the error incurred during step~(\ref{step v}) remains. A similar argument applies for the components corresponding to $\ket{10}_{q}$ and $\ket{11}_{q}$, resulting in an improved fidelity in~(\ref{eq:overlap with Kerr step x}) as compared to (\ref{eq:overlap with Kerr step vi}). Nonetheless, the relatively large influence of the Kerr effect in~(\ref{eq:overlap with Kerr step iv})-(\ref{eq:overlap with Kerr step vi}) is problematic, and indeed does not justify terminating the power series after order $\epsilon^{2}$. We shall hence modify our protocol in a similar fashion as discussed in~\cite{LeghtasEtAl2013} to correct for the leading order Kerr effect altogether.

\hypertarget{sec:corrected 2 qubit protocol2}{}
\subsection{Corrected two-qubit protocol}\label{sec:corrected 2 qubit protocol}
\bookmark[named=sec:corrected 2 qubit protocol,level=0,dest=sec:corrected 2 qubit protocol2]{III. C. Corrected two-qubit protocol}

\begin{figure*}[ht!]
    \hspace*{-4.5mm}\includegraphics[width=0.47\textwidth]{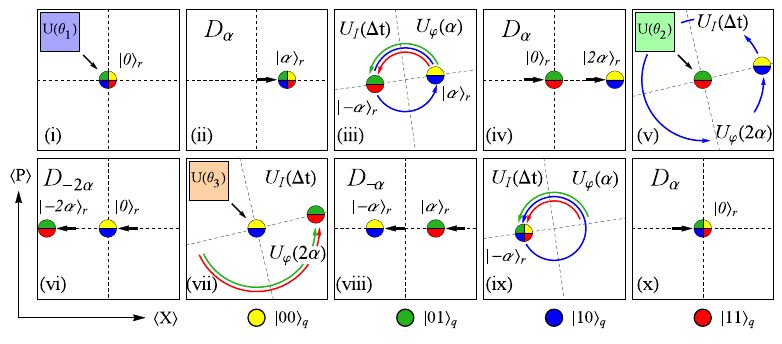}
    \hspace*{1mm}
    \includegraphics[width=0.49\textwidth]{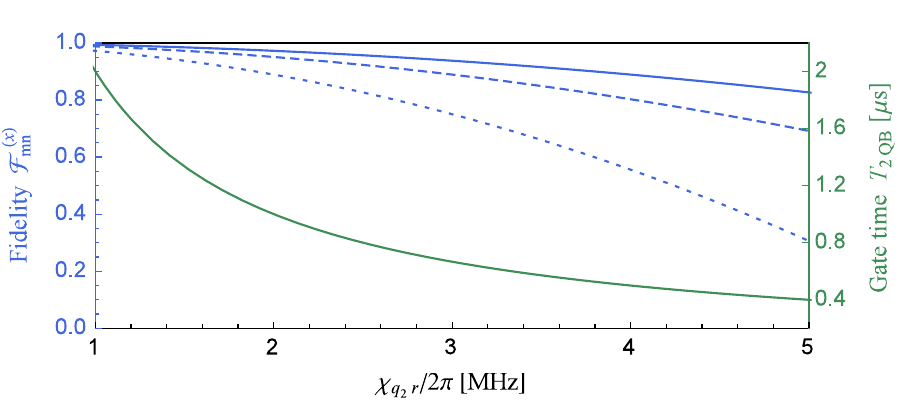}\\
    \vspace*{-7mm}
    \flushleft{\textbf{(a)}}\hspace*{0.49\textwidth}\textbf{(b)}\\
    \hspace*{2mm}\includegraphics[width=0.5\textwidth,trim={6cm 0 7cm 0},clip]{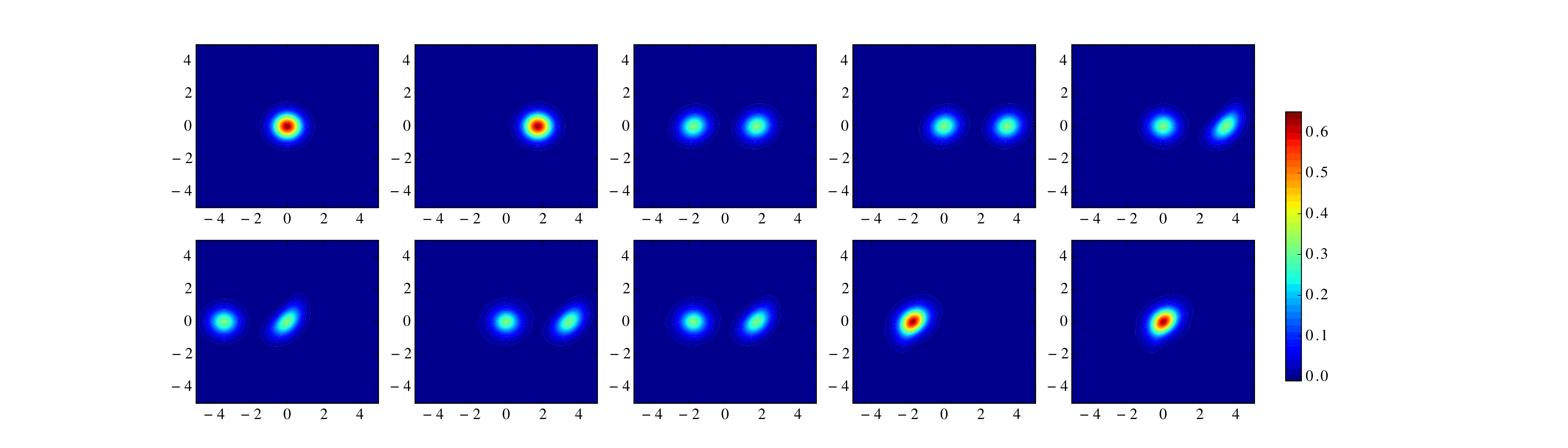}
    \includegraphics[width=0.45\textwidth,trim={5cm 0 7cm 0},clip]{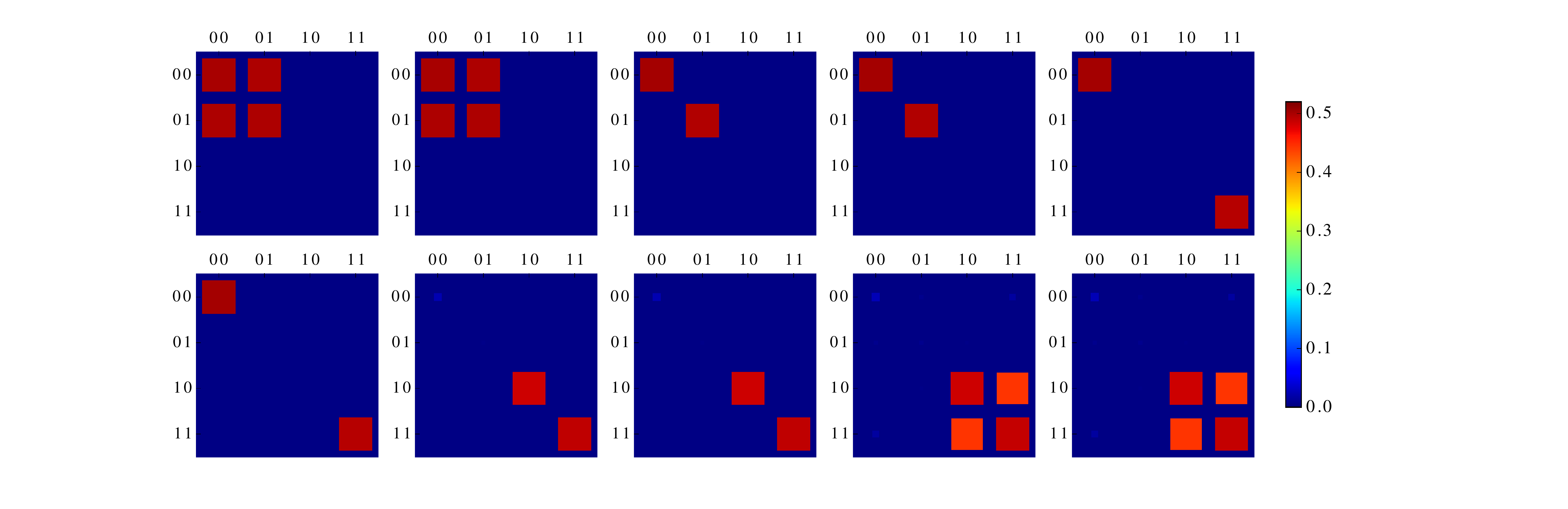}\\
    \vspace*{-7mm}
    \flushleft{\textbf{(c)}}\hspace*{0.49\textwidth}\textbf{(d)}
    \caption{\label{fig:coherent controlization Kerr corrected}
    \textbf{Corrected two-qubit protocol}: \textbf{(a)} The ideal two-qubit protocol from Fig.~\ref{fig:coherent controlization circuit diagram} is modified to correct for the leading order cavity Kerr effect. After each period of free time evolution, the basis in the phase space is adjusted by an angle $\varphi_{\alpha}=\epsilon(2|\alpha|^{2}+1)$ [steps~(\ref{step iii}) and~(\ref{step ix})] or $\varphi_{2\alpha}=\epsilon(8|\alpha|^{2}+1)$ [steps~(\ref{step v}) and~(\ref{step vii})], corresponding to the largest displacement in the previous step. \textbf{(b)} The gate fidelity (top three blue curves) as measured by $\Fmnk{mn}{x}$ (the overlap of the final cavity states with the vacuum) and the gate time $T_{2QB}=4\Delta t$ (in $\mu$s, bottom green) are shown as functions of $\chi_{q_{2}r}/(2\pi)=\Delta\omega/(2\pi)$ in MHz. The fidelities (not taking into account decoherence) are shown for values $\chi_{qq}/(2\pi)=300$~MHz, and average photon numbers $\bar{n}=1.5$ (solid, top), $\bar{n}=2$ (dashed, second from the top), and $\bar{n}=3$ (dotted, third from the top). \textbf{(c)} A simulation (using the QuTiP library~\cite{JohanssonNationNori:QuTip2:2013}) of the Wigner function of the reduced cavity state in the corrected protocol steps (\ref{step i})-(\ref{step x}) from~(\textbf{a}) is shown for $\bar{n}=3$, $\chi_{q_{2}r}/(2\pi)=1.5$~MHz, and $\chi_{qq}/(2\pi)=300$~MHz [corresponding to the third curve from the top in \textbf{(b)}] when dephasing and amplitude damping for the qubits (with coherence times $\tau_{\phi}=\tau_{q}=100~\mu$s), and photon loss for the resonator ($\tau_{r}=100~\mu$s) are taken into account. Note that even without loss the reduced cavity state $\sum_{m,n}|c_{mn}^{\protect\raisebox{-0.5pt}{\tiny{(k)}}}|^{2}
    \psistepij{mn}{k}\!\!\psistepijbra{mn}{k}$ does not feature interference fringes. In addition, the application of the conditional qubit unitaries can decrease the interference between the vacuum and excited state components in the reduced cavity state, partially restoring the rotational symmetry of the Wigner function. This can be seen, e.g., in the transition from step~(\ref{step vi}) to~(\ref{step vii}) in \textbf{(c)}. The simulated protocol, for which the reduced qubit state amplitudes $|_{q\!\!}\bra{\!\mu\nu\!}\bigl(\tr_{r}\rho\bigr)\ket{\!mn\!}_{\!q}\!|$ $(m,n,\mu,\nu=0,1)$ are shown in \textbf{(d)}, was run for the initial qubit state $\ket{00}_{\!q}$, and with rotation angles $2\theta_{1}=\theta_{2}=\theta_{3}=\pi$. Including decoherence, the simulation with $\bar{n}=3$, $\chi_{q_{2}r}/(2\pi)=1.5$~MHz, and $\chi_{qq}/(2\pi)=300$~MHz yielded  $\Fmnk{mn}{x}=95\%$ and the overlap of the final state of the qubits with the target state amounted to~$93\%$. Further details on this, and additional simulations with other choices of parameters can be found in the \hyperref[sec:appendix]{Appendix}.}
\end{figure*}

To completely compensate for the contribution of the linear order Kerr effect, the displacements after each period of free time evolution are adjusted by angles $\varphi_{\gamma}=\epsilon(2|\gamma|^{2}+1)$, where $\gamma$ is the maximal displacement of the coherent state components during the time evolution. Alternatively, this may be seen as a suitable change of basis in the phase space, as illustrated in Fig.~\ref{fig:coherent controlization Kerr corrected}~(a).

For the evaluation of the fidelities, we hence include the transformations $U_{\varphi}(\gamma)=\exp(-i\varphi_{\gamma}\boldsymbol{a}^{\hspace*{-0.5pt}\dagger}\!\boldsymbol{a})$, that is
\begin{subequations}
\label{eq:resonator states with Kerr corrected}
\begin{align}
    \psistepij{mn}{iv}\ket{\hspace*{-0.8pt}mn\!}_{\!q}    &=
    D_{\alpha}U_{\!\varphi}(\alpha)U_{I\hspace*{-0.5pt},K}(\Delta t)D_{\alpha}\ket{0}_{\hspace*{-0.8pt}r}\!\ket{\hspace*{-0.8pt}mn\!}_{\!q},
    \label{eq:resonator states with Kerr corrected step v}\\[1mm]
    \psistepij{mn}{vi}\ket{\hspace*{-0.8pt}mn\!}_{\!q}    &=D_{-2\alpha}U_{\!\varphi}(2\alpha)U_{I\hspace*{-0.5pt},K}(\Delta t)\psistepij{mn}{iv}\ket{\hspace*{-0.8pt}mn\!}_{\!q} ,
    \label{eq:resonator states with Kerr corrected step vii}\\[1mm]
    \psistepij{mn}{x}\ket{\hspace*{-0.8pt}mn\!}_{\!q}     &=
    D_{\alpha}U_{\!\varphi}(\alpha)U_{I\hspace*{-0.5pt},K}(\Delta t)D_{-\alpha}U_{\!\varphi}(2\alpha)\nonumber\\
    &\ \ \ \times U_{I\hspace*{-0.5pt},K}(\Delta t)\psistepij{mn}{vi}\ket{\hspace*{-0.8pt}mn\!}_{\!q}.
    \label{eq:resonator states with Kerr corrected step x}
\end{align}
\end{subequations}
Proceeding as before, we arrive at the corrected fidelities for the overlap with the vacuum after steps~(\ref{step iv}), (\ref{step vi}), and (\ref{step x}), given by
\begin{subequations}
\label{eq:corrected fidelities}
\begin{align}
    \Fmnk{01}{iv}&=\,\Fmnk{11}{iv}\,=\,1\,-\,2\bar{n}^{2}\epsilon^{2}\,+\,\mathcal{O}(\epsilon^{3})\,,
    \label{eq:corrected fidelities step iv}\\[1mm]
    \Fmnk{00}{vi}&=\,\Fmnk{10}{vi}\,=\,1\,-\,50\bar{n}^{2}\epsilon^{2}\,+\,\mathcal{O}(\epsilon^{3})\,,
    \label{eq:corrected fidelities step iv}\\[1mm]
    \Fmnk{mn}{x}&=\,1\,-\,72\bar{n}^{2}\epsilon^{2}\,+\,\mathcal{O}(\epsilon^{3})\,.
\end{align}
\end{subequations}

The fidelity of the protocol as quantified by $\Fmnk{mn}{x}$ (not yet taking into account decoherence) hence also depends on the displacement via the average photon number $\bar{n}=|\alpha|^{2}$. Crucially, these values have to be chosen such that the overlaps between the different coherent state components remain small in steps~(\ref{step v}) and~(\ref{step vii}). For values $\bar{n}=1.5$ and $2$ one finds $|\!\scpr{\hspace*{-0.5pt}0}{\hspace*{-0.5pt}2\alpha\hspace*{-0.5pt}}\!|^{2}=\exp(-4|\alpha|^{2})\approx2.5\times10^{-3}$ and $3.3\times10^{-4}$, respectively. For $\chi_{qq}/(2\pi)=300$~MHz and $\bar{n}=1.5$, one may reach fidelities $\Fmnk{mn}{x}$ between $99\%$ and $94\%$ for qubit-cavity cross-Kerr coefficients ranging from $\chi_{q_{2}r}/(2\pi)=1$~MHz to $3$~MHz, see Fig.~\ref{fig:coherent controlization Kerr corrected}~(b). In the \hyperref[sec:appendix]{Appendix} we further include decoherence effects \textemdash\ dephasing and amplitude damping of the qubits with coherence times $\tau_{\phi}=30-100~\mu$s and $\tau_{q}=20-100~\mu$s, respectively, as well as photon loss in the cavity with coherence times $\tau_{r}=100~\mu$s, but we consider the single-qubit operations to be perfect. Employing simulations coded in PYTHON using the QuTiP library~\cite{JohanssonNationNori:QuTip2:2013} we find that fidelities of $95\%$ are reasonably achievable, e.g., for $\bar{n}=3$, $\chi_{q_{2}r}/(2\pi)=1.5$~MHz, $\chi_{qq}/(2\pi)=300$~MHz and coherence times of $100~\mu$s for a range of initial states and angles $\theta_{i}\ (i=1,2,3)$, e.g., as shown in Fig.~\ref{fig:coherent controlization Kerr corrected}~(c,d).

\hypertarget{sec:other corrections2}{}
\subsection{Other corrections}\label{sec:other corrections}
\bookmark[named=sec:other corrections,level=0,dest=sec:other corrections2]{III. D. Other corrections}

Before we finally turn to the extension to three (and more) qubits, let us remark on additional possible sources of decreases in fidelity. In~a similar way, in which the cavity inherits the Kerr term from the anharmonicities of the qubits, the qubits are also coupled via the cavity, even if direct capacitive coupling can be avoided by arranging the qubits to be spatially well-separated. The corresponding term of the form $\chi_{q_{1}q_{2}}\bdn{1}\bn{1}\bdn{2}\bn{2}$ leads to additional phases that are acquired by the components $\psistepij{11}{k}\ket{11}_{q}$ during the time evolution. This effect is at most of a~size comparable to the cavity Kerr effect, which is required to be small to achieve high-fidelities.

However, when the effect becomes non-negligible, it can be corrected by applying an appropriate phase gate along with $U(\theta_{2})$ during step~(\ref{step v}). In a regime where this becomes necessary, additional measures are further required to compensate the cavity Kerr effect beyond the linear order corrections that we have considered so far. If required, this can be achieved by a scheme, recently proposed in Ref.~\cite{HeeresVlastakisHollandKrastanovAlbertFrunzioJiangSchoelkopf2015}, that relies on additional ancilla qubits. In the strong dispersive regime, the frequency shift of the ancilla qubit due to different photon numbers in the cavity is used for photon-number selective phase gates that compensate the phases acquired by the different Fock state components of the coherent states. When such corrections are used to eliminate the cavity self-Kerr effect, the fidelity $\Fmnk{mn}{x}$ of our protocol with parameters as in Fig.~\ref{fig:coherent controlization Kerr corrected} can reach $99\%$.

Another potential source of errors lies in imperfections in the circuit fabrication that may cause~a small deviation~$\epsilon$ from the desired ratio $\chi_{q_{2}r}=\frac{\chi_{q_{1}r}}{2}$ of the qubit-cavity cross-Kerr coefficients, that is, one may encounter a situation where $\chi_{q_{2}r}=\frac{\chi_{q_{1}r}}{2}(1+\delta)$ for some small (real) $|\delta|\ll1$. This causes an additional rotation for the coherent state components corresponding to $\ket{01}_{q}$ and $\ket{11}_{q}$ in steps~(\ref{step iii}), (\ref{step vii}) and (\ref{step ix}), but not for the components corresponding to $\ket{00}_{q}$ and $\ket{10}_{q}$. For steps~(\ref{step iii}) and (\ref{step vii}) this effect can be compensated by modified displacements in steps~(\ref{step iv}), (\ref{step vi}) and (\ref{step viii}). However, to recombine all coherent state components in step~(\ref{step ix}), an ``echo"-type operation is required. That is, step~(\ref{step ix}) is amended in the following way: after a waiting time $t\pr=\frac{\pi}{\chi_{q_{1}r}}(\frac{2+\delta}{1+\delta})<\Delta t$ an unconditional $\pi$-pulse~\cite{LeghtasEtAl2013} is applied to the second qubit, which exchanges the subspaces $\{\ket{00}_{q},\ket{10}_{q}\}$ and $\{\ket{01}_{q},\ket{11}_{q}\}$. After another waiting period of duration $t\prpr=\frac{2\pi}{\chi_{q_{1}r}}-t\pr$ a final $\pi$-pulse to the second qubit restores the original qubit state. Step~(\ref{step x}) can then be executed with an appropriately modified displacement to complete the protocol. With the possibility for these corrections in mind, we now turn to the three-qubit protocol.

\begin{figure*}[ht!]
    \textbf{(a)}\includegraphics[width=0.89\textwidth]{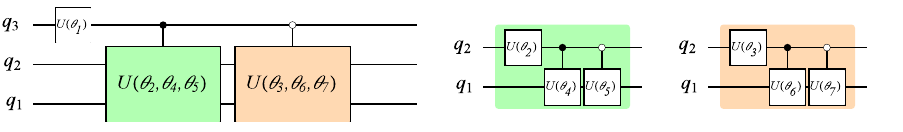}
    \textbf{(b)}\includegraphics[width=0.89\textwidth]{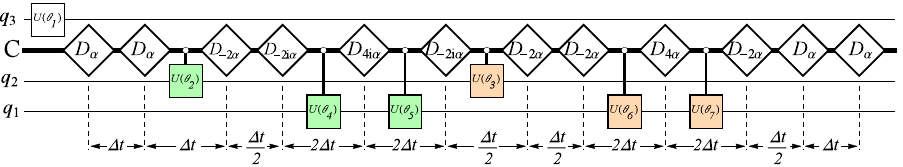}
    \includegraphics[width=0.98\textwidth]{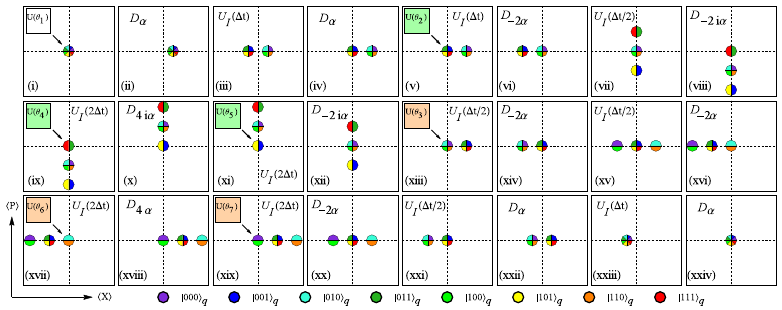}\\
    \vspace*{-6mm}
    \flushleft{\textbf{(c)}}
    \caption{\label{fig:coherent controlization three qubits}
    \textbf{Three-qubit protocol for coherent controlization}: The circuit shown in \textbf{(a)}, where the filled dots ``$\bullet$" on the controlled operations indicate that the unitaries on the target are conditioned on the control qubit state~$\ket{1}_{q_{3}}$, while the hollow dots ``$\circ$" denote conditioning on the control qubit state~$\ket{0}_{q_{3}}$, represents~a three-qubit probability unitary $U_{j}$. \textbf{(b)} shows an extended circuit diagram that realizes the circuit in~(a) by adding control to the single-qubit $Y$-rotations $U(\theta_{i})$. The uppermost, and the two lowest horizontal lines indicate qubits~$3$, $2$ and~$1$, respectively, while the second line from the top represents the cavity mode. The white, diamond-shaped rectangles represent unconditional displacements of all cavity modes, and the control lines between the cavity and the single-qubit operations indicate operations conditioned on the resonator mode being in the ground state. The waiting time between the displacements varies between $\Delta t/2=\pi/(2\Delta\omega)$ and the full period $2\Delta t$. \textbf{(c)} shows the phase space representation of the coherent state components $\psistepij{ijk}{l}$\ $(i,j,k=0,1)$ corresponding to the qubit states $\ket{ijk}_{q}$ for the steps $l=\mathrm{(i)},\ldots,\mathrm{(xxiv)}$ of the ideal three-qubit protocol.}
\end{figure*}

\section{Coherent controlization for three qubits \& scalability}\label{sec:scaling}
\bookmark[named=sec:scaling,level=-1,dest=sec:scaling2]{IV. Coherent controlization for three qubits \& scalability}
\vspace*{-15mm}
\hypertarget{sec:scaling2}{}
\vspace*{15mm}

Let us now discuss the extension of our protocol to $3$ qubits and beyond. We will consider the ideal-three qubit protocol for the realization of~a probability unitary $U_{j}$ as a proof-of-principle example from which the construction of the protocol for any number of qubits may be inferred. To extend the previous protocol, first note that the cross-Kerr interaction $H_{\hspace*{-0.5pt}I}/\hbar=-\,\sum_{i}\chi_{q_{i}r}\,\boldsymbol{a}^{\dagger}\boldsymbol{a}\,\bdn{i}\bn{i}$ again splits the resonator frequency into spectral lines for all the qubit states. The sidebands are equally spaced by~$\Delta\omega$ by selecting $\chi_{q_{1}r}/4=\chi_{q_{2}r}/2=\chi_{q_{3}r}=\Delta\omega$. The ideal protocol, detailed in Fig.~\ref{fig:coherent controlization three qubits}, then relies on the same type of operations as before, \emph{unconditional displacements} (now with amplitudes up to $4\alpha$), and \emph{waiting periods} (now of durations $\Delta t/2,\Delta t$, and $2\Delta t$), and qubit operations conditioned on the cavity being in the vacuum state.

Increasing the number of qubits hence brings about exponential scaling of several quantities of interest \textemdash\ as expected when exponentially increasing the dimension of the state space that we wish to explore. To separate the coherent state components corresponding to all different subspaces, larger displacements and finer graining of the rotation angles in phase space are required. That is, in addition to rotations by half-periods, for three qubits also quarter periods are necessary, and for $n$-qubits, rotations by $2\pi/2^{n-1}$ are needed. In addition, waiting periods $2\Delta t$ corresponding to the full period become obligatory when applying the conditional qubit rotations. Starting from an ideal $n$ qubit protocol (for $n\geq3$), an additional qubit may be added by inserting $2^{n-1}$ sequences of operations between each pair of conditioned unitaries on the qubit previously labelled $q_{1}$ (e.g., $U(\theta_{4})$ and $U(\theta_{5})$ for $n=3$). Each sequence consists of (at most) $4$ displacements and two conditioned unitaries on the new qubit and the waiting periods after the $4$ new displacements are of duration $\Delta t/2^{n-1}$, $2\Delta t$, $2\Delta t$ and $2\Delta t(1-2^{-n})$, respectively. We hence find that, in an ideal $n$-qubit protocol, the number of displacements is $N_{D}(n\geq3)=2^{n+1}-4$, the number of qubit unitaries is $N_{U}(n)=2^{n}-1$, and the overall duration is $T(n\geq3)=(3\times2^{n}-11)\Delta t$. In a nonideal protocol, correction operations have to be included as well. For instance, for each conditioned qubit unitary $2$ echo pulses (as discussed in Section~\ref{sec:other corrections}) may need to be added, increasing the total number of qubit operations to $3\times 2^{n-1}$, but leaving the gate time unchanged. The exponential increase in dimension hence carries over to the scaling of the number of conditional qubit operations, and to the overall gate time. Since increases in gate times, displacements and qubit-cavity cross-Kerr coefficients $\chi_{q_{i}r}$ all lead to increased disturbance due to the cavity Kerr effect, its compensation using photon-number selective phase gates~\cite{HeeresVlastakisHollandKrastanovAlbertFrunzioJiangSchoelkopf2015} becomes ubiquitous, despite the possibility to correct for the linear order Kerr effect in Eq.~(\ref{eq:linear Kerr effect}) even in the presence of three individual phase space components with different displacements.

Nonetheless, it should be mentioned that such challenges for the scaling of quantum computational architectures are expected for every physical platform, and are hence not specific to our proposal. For our three-qubit protocol, simulations (using the QuTiP library~\cite{JohanssonNationNori:QuTip2:2013}) that take into account partial correction of the linear Kerr effect and decoherence yield fidelities for the cavity state of $80\%$ ($64\%$ for the qubits) for the system parameters $\chi_{q_{3}r}/(2\pi)=0.3$~MHz, $\chi_{qq}/(2\pi)=300$~MHz and $\bar{n}=1$, see the \hyperref[sec:appendix]{Appendix}. However, when corrections as discussed in~\cite{HeeresVlastakisHollandKrastanovAlbertFrunzioJiangSchoelkopf2015} are included, the same parameters yield fidelities of up to $95\%$ for the cavity, and $75\%$ for the qubits.

\vspace*{-55mm}
\hypertarget{sec:discussion2}{}
\vspace*{55mm}
\section{Discussion}\label{sec:discussion}
\bookmark[named=sec:discussion,level=-1,dest=sec:discussion2]{V. Discussion}
\vspace*{-1mm}

We have introduced a protocol for coherent controlization, that is, adding control to (a set of) unspecified or unknown unitaries, using superconducting qubits coupled to a microwave resonator. This task is of interest for the flexible realization of quantum computational architectures, but also of great importance for the adaptiveness of quantum-enhanced learning agents~\cite{DunjkoFriisBriegel2015}. We have selected an example from the latter context, the reflective projective simulation model~\cite{PaparoDunjkoMakmalMartinDelgadoBriegel2014} for artificial intelligence, where coherent controlization is already useful at the lowest level of the deliberation algorithm to construct coherent encodings of Markov chains \textemdash\, the probability unitaries.

We have given explicit protocols for the realization of these unitary operations using transmon qubits~\cite{LeghtasEtAl2013}. Within the strong dispersive regime, we exploit the coupling between the qubits and the resonator. The cavity mode here acts as a bus between the qubits, playing the role of the additional degree of freedom necessary for adding control to unknown unitaries, similar, e.g., to the vibrational modes of trapped ions~\cite{FriisDunjkoDuerBriegel2014}. We have provided a detailed discussion of the role of the cavity Kerr effect, the strongest source of disturbance, in our protocol, including corrections for the linear order effect. Based on these considerations, and bolstered by numerical simulation including photon loss, as well as amplitude- and phase-damping for the qubits (featured in the \hyperref[sec:appendix]{Appendix}), we conclude that a possible experimental realization of our protocol with two qubits may achieve high fidelities (up to $95\%$ for the cavity and $93\%$ for the qubits) for reasonable ranges of the system parameters. We hence consider our proposal for two qubits to be readily implementable using current superconducting technology. For three (or more) qubits, an implementation is still possible, although the significant drop in fidelity ($<80\%$ and $<64\%$ for the cavity and qubits respectively) suggests that upscaled versions of our protocol may require additional corrections of the Kerr effect~\cite{HeeresVlastakisHollandKrastanovAlbertFrunzioJiangSchoelkopf2015}, which can bring the fidelities up to $95\%$ (cavity) and $75\%$ (qubits).\\
\vspace*{-5mm}

\hypertarget{acknowledgements}{}
\begin{acknowledgements}
\vspace*{-4mm}
\bookmark[named=acknowledgements,level=-1,dest=acknowledgements]{Acknowledgements}
We are grateful to Vedran Dunjko, Oscar Gargiulo, and Steve Girvin for valuable discussions and comments. N.~F. thanks Yale University for hospitality. This work was supported by the Austrian Science Fund (FWF) through Grant No.~SFB FoQuS F4012, and by the Templeton World Charity Foundation through Grant No.~TWCF0078/AB46.\\

\vspace*{5.5mm}
\end{acknowledgements}


\vspace*{-9mm}
\hypertarget{references}{}




\appendix*
\hypertarget{appendix}{}
\section*{Appendix:\\[1mm] Supplementary Information}\label{sec:appendix}
\bookmark[named=appendix,level=-2,dest=appendix]{Appendix: Supplementary Information}

\renewcommand\appendixname{}
\renewcommand{\thesection}{A.\arabic{section}}
\setcounter{figure}{0}
\renewcommand{\thefigure}{\Alph{section}.\arabic{figure}}

The Appendix is structured as follows. In \ref{sec:coupling hamiltonian} we provide a derivation of the Hamiltonian of the transmon qubits coupled to a resonator. The aim of this section is to give a transparent account of all approximations made in the system we consider and to give an introduction to this kind of superconducting qubits that is accessible for non-specialists. In \ref{sec:simulations} we then present the details of the simulations that were conducted to assess the robustness of our protocol in the presence of the Kerr effect, dephasing and amplitude damping of the qubits, and photon loss in the resonator.

\renewcommand{\thesection}{Appendix \Alph{section}}

\hypertarget{sec:coupling hamiltonian2}{}
\section{Transmon qubits coupled to a microwave resonator}\label{sec:coupling hamiltonian}
\bookmark[named=sec:coupling hamiltonian,level=-1,dest=sec:coupling hamiltonian2]{Appendix A: Transmon qubits coupled to a microwave resonator}

\renewcommand{\thesubsection}{\Alph{section}.\Roman{subsection}}

\renewcommand{\theequation}{\Alph{section}.\arabic{equation}}

\hypertarget{sec:josephson junctions to transmons2}{}
\subsection{From Josephson junctions to transmons}\label{sec:josephson junctions to transmons}
\bookmark[named=sec:josephson junctions to transmons,level=0,dest=sec:josephson junctions to transmons2]{Appendix A. I. From Josephson junctions to transmons}

A superconducting $LC$\textendash circuit may be thought of as~a harmonic oscillator, where the position and conjugate momentum variables are the flux~$\boldsymbol{\Phi}$ through the inductor and the charge~$\boldsymbol{Q}$ on the capacitor plates, respectively. Indeed, these quantities have to be treated as operators satisfying the canonical commutation relation~$\comm{\boldsymbol{\Phi}}{\boldsymbol{Q}}=i\hbar$, see, e.g., Ref.~\cite{DevoretWallraffMartinis2004}. In other words, the system Hamiltonian $H=\boldsymbol{Q}^{2}/(2C)+\boldsymbol{\Phi}^{2}/(2L)$ may be written as
\begin{align}
    H   &=\,\hbar\omega(\boldsymbol{q}^{\dagger}\boldsymbol{q}+1/2)
\end{align}
in terms of the ladder operators defined by
\begin{subequations}
\label{eq:LC ladder operators}
\begin{align}
    \boldsymbol{Q} &=\,-i\,\sqrt{\frac{\hbar}{2Z}}\left(\boldsymbol{q}-\boldsymbol{q}^{\dagger}\right)\,,
    \label{eq:LC Q operator}\\[1mm]
    \boldsymbol{\Phi} &=\, \sqrt{\frac{\hbar Z}{2}}\left(\boldsymbol{q}+\boldsymbol{q}^{\dagger}\right)\,,
    \label{eq:LC Phi operator}
\end{align}
\end{subequations}
with $\omega=1/\sqrt{LC}$, $Z=\sqrt{L/C}$, and $\left[\boldsymbol{q},\boldsymbol{q}^{\dagger}\right]=1$. The introduction of a~Josephson junction (a~thin insulating barrier separating two pieces of superconducting material) creates~a non-linearity in the system that allows identifying two energy levels as the qubit levels. The Hamiltonian is then modified to
\begin{align}
  H &= \frac{\boldsymbol{Q}^2}{2C_{\Sigma}} - \frac{\Phi_{0}^{2}}{L_{J}}\cos\left(\frac{\boldsymbol{\Phi}}{\Phi_{0}}\right)\,,
  \label{eq:Josephson junction circuit Hamiltonian}
\end{align}
where the junction inductance $L_{J}$ is typically expressed via the Josephson energy $E_{J}$, that is, $L_{J}=\Phi_{0}^{2}/E_{J}$, with the magnetic flux quantum $\Phi_{0}=\hbar/(2e)$, and $\boldsymbol{\Phi}/\Phi_{0}$ is the phase difference across the junction. The original capacitance and the Josephson junction now form~a superconducting island \textemdash\ a Cooper-pair box (CPB) \textemdash\  with capacitance $C_{\Sigma}=C+C_{J}$, which can be written in terms of the charging energy $E_{C}=e^{2}/(2C_{\Sigma})$. The charge in the CPB depends on $\boldsymbol{n}$, the number of transferred Cooper pairs (with charge $2e$), and the effective offset charge $2e\hspace*{0.5pt}n_{o}$, i.e., $\boldsymbol{Q}/(2e)=\boldsymbol{n}-n_{o}$. With the replacement $Z\rightarrow Z_{J}=\sqrt{L_{J}/C_{\Sigma}}$ one may define raising and lowering operators in full analogy to~(\ref{eq:LC ladder operators}). To remove the strong dependence of the energy-level splitting on the (environmentally induced) offset charge, an additional shunting capacitor with large capacitance $C_{B}$ can be inserted into the circuit in parallel with the Josephson junction to create the \emph{transmon}~\cite{Koch-Schoelkopf2007}. This strongly increases $C_{\Sigma}\rightarrow C_{\Sigma}=C_{B}+C+C_{J}$, such that $E_{C}/E_{J}\ll1$. Inserting the ladder operators $\boldsymbol{q}$ and $\boldsymbol{q}^{\dagger}$ (with $Z\rightarrow Z_{J}$) into Eq.~(\ref{eq:Josephson junction circuit Hamiltonian}) and expanding in powers of $\sqrt{E_{C}/E_{J}}$ we arrive at
\begin{align}
    H &=\,\sqrt{2E_{J}E_{C}}-E_{J}-\frac{E_{C}}{4}\,+\,
    \left(\sqrt{8E_{J}E_{C}}-E_{C}\right)\boldsymbol{q}^{\dagger}\boldsymbol{q}
    \nonumber\\[1mm]
    &\  - \frac{E_C}{3}\left(\boldsymbol{q}^\dag\boldsymbol{q}^\dag\boldsymbol{q}^\dag\boldsymbol{q} + \boldsymbol{q}^\dag\boldsymbol{q}\boldsymbol{q}\boldsymbol{q}\right) - \frac{E_C}{12}\left(\boldsymbol{q}\boldsymbol{q}\boldsymbol{q}\boldsymbol{q} + \boldsymbol{q}^\dag\boldsymbol{q}^\dag\boldsymbol{q}^\dag\boldsymbol{q}^\dag\right)\nonumber\\[1mm]
    &\ - \frac{E_C}{2}\left(\boldsymbol{q}\boldsymbol{q} + \boldsymbol{q}^\dag\boldsymbol{q}^\dag + \boldsymbol{q}^\dag\boldsymbol{q}^\dag\boldsymbol{q}\boldsymbol{q}\right).
    \label{eq:transmon full Hamiltonian}
\end{align}
Within the rotating wave approximation, the Hamiltonian of Eq.~(\ref{eq:transmon full Hamiltonian}) can be truncated to the effective transmon qubit Hamiltonian
\begin{align}
  H_{q} &= \hbar\omega_{Q}\,\boldsymbol{q}^{\dagger}\boldsymbol{q}\,-\,\frac{\hbar\beta}{2}\boldsymbol{q}^\dag\boldsymbol{q}^\dag\boldsymbol{q}\boldsymbol{q}\,,
  \label{eq:transmon qubit Hamiltonian}
\end{align}
where $\omega_{Q}\equiv\left(\sqrt{8E_{J}E_{C}}-E_{C}\right)/\hbar$ is the transmon qubit frequency and $\beta\equiv E_{C}/\hbar$ is the anharmonicity.

\hypertarget{sec:Transmons Coupled to a Resonator2}{}
\subsection{Transmons coupled to a resonator}\label{sec:Transmons Coupled to a Resonator}
\bookmark[named=sec:Transmons Coupled to a Resonator,level=0,dest=sec:Transmons Coupled to a Resonator2]{Appendix A. II. Transmons coupled to a resonator}

Next, we are interested in describing~a transmon qubit that is capacitively coupled to~a microwave resonator. Within the rotating wave approximation the joint system is described by the Hamiltonian
\begin{align}
  H_{qr}/\hbar &=\, \omega_R\boldsymbol{c}^\dag\boldsymbol{c} + \omega_Q\boldsymbol{q}^\dag\boldsymbol{q} + g\left(\boldsymbol{c}^\dag\boldsymbol{q} + \boldsymbol{c}\boldsymbol{q}^\dag\right) - \frac{\beta}{2}\boldsymbol{q}^\dag\boldsymbol{q}^\dag\boldsymbol{q}\boldsymbol{q}\,,
  \label{eq:single transmon coupled to resonator rot wave approx}
\end{align}
where $g$ is the qubit-cavity coupling, $\omega_R$ and $\omega_Q$ are the frequencies of the isolated qubit and cavity, respectively, and the operators $\boldsymbol{c}$ and $\boldsymbol{c}^{\dagger}$ are the resonator ladder operators satisfying $\comm{\boldsymbol{c}}{\boldsymbol{c}^{\dagger}}=1$. The terms in~$H_{qr}$ that are quadratic in the mode operators can be diagonalized by~a Bogoliubov transformation. That is, one introduces the dressed operators $\boldsymbol{a}$ and $\boldsymbol{b}$, such that
\begin{align}
 \boldsymbol{c} & = \cos{\theta}~\boldsymbol{a} + \sin{\theta}~\boldsymbol{b}\,, \nonumber\\
 \boldsymbol{q} & = -\sin{\theta}~\boldsymbol{a} + \cos{\theta}~\boldsymbol{b}\,,
\end{align}
with $\tan{2\theta} = 2g/\Delta$, where $\Delta \equiv \omega_Q - \omega_R$ is the qubit-cavity detuning. The excitations of the dressed modes can no longer be uniquely attributed to just the resonator or just the qubit. However, in the strong dispersive regime, where $g/\Delta\ll1$, the mixing angle is small, $\theta\approx g/\Delta$ and excitations created by $\boldsymbol{c}^{\dagger}$ ($\boldsymbol{q}^{\dagger}$) are associated ``mostly" with the cavity (qubit). The Hamiltonian transforms to
\begin{align}
H_{qr}/\hbar & =\, \tilde{\omega}_{r}\boldsymbol{a}^{\dagger}\boldsymbol{a}\,+\,\tilde{\omega}_{q}\boldsymbol{b}^{\dagger}\boldsymbol{b}
    \,-\,2\beta\sin^{2}\!\theta\cos^{2}\!\theta~\boldsymbol{a}^{\dagger}\boldsymbol{a}\boldsymbol{b}^{\dagger}\boldsymbol{b}
    \nonumber\\[1mm]
    &\ -\,\frac{\beta}{2}\cos^{4}\!\theta~\boldsymbol{b}^{\dagger}\boldsymbol{b}^{\dagger}\boldsymbol{b}\boldsymbol{b}\,-\,
    \frac{\beta}{2}\sin^{4}\!\theta\boldsymbol{a}^{\dagger}\boldsymbol{a}^{\dagger}\boldsymbol{a}\boldsymbol{a}
    \nonumber\\[1mm]
    &\ +\,\beta\sin\theta\cos\theta~\Bigl[\sin^{2}\!\theta\bigl(\boldsymbol{a}^{\dagger}\boldsymbol{a}^{\dagger}\boldsymbol{a}\boldsymbol{b}
        +\,\boldsymbol{a}^{\dagger}\boldsymbol{a}\boldsymbol{a}\boldsymbol{b}^{\dagger}\bigr)
    \nonumber\\[1mm]
    &\ \ +\,\cos^{2}\!\theta~\bigl(\boldsymbol{a}^{\dagger}\boldsymbol{b}^{\dagger}\boldsymbol{b}\boldsymbol{b}
        +\,\boldsymbol{a}\boldsymbol{b}^{\dagger}\boldsymbol{b}^{\dagger}\boldsymbol{b}\bigr)
    \nonumber\\[1mm]
    &\ \ +\,\sin\theta\cos\theta~\bigl(\boldsymbol{a}^{\dagger}\boldsymbol{a}^{\dagger}\boldsymbol{b}\boldsymbol{b}
        +\,\boldsymbol{a}\boldsymbol{a}\boldsymbol{b}^{\dagger}\boldsymbol{b}^{\dagger}\bigr)\Bigr]\,,
\end{align}
where the dressed mode frequencies are given by
\begin{subequations}
\label{eq:dressed mode frequencies}
\begin{align}
    \tilde{\omega}_{r}  &=\,\omega_{R}\cos^{2}\!\theta\,+\,\omega_{Q}\sin^{2}\!\theta\,-\,g\sin(2\theta)\,,
    \\[1mm]
    \tilde{\omega}_{q}  &=\,\omega_{R}\sin^{2}\!\theta\,+\,\omega_{Q}\cos^{2}\!\theta\,+\,g\sin(2\theta)\,.
\end{align}
\end{subequations}
The terms in the square bracket can be seen to oscillate rapidly, and we may therefore remove these terms in another rotating wave approximation. With the notation $\chi_{qr}\equiv(\beta/2)\sin^{2}\!(2\theta)$, $\chi_{qq}\equiv\beta\cos^{4}\!\theta$, and
\begin{align}
    \chi_{rr}   &\equiv\,\frac{\chi_{qr}^{2}}{4\chi_{qq}}\,=\,\beta\sin^{4}\!\theta\,,
    \label{eq:chi rr one qubit}
\end{align}
we then arrive at the Hamiltonian of~a single transmon qubit coupled to~a resonator in the dispersive limit
\begin{align}
H_{qr}/\hbar    &=\,\tilde{\omega}_{r}\boldsymbol{a}^{\dagger}\boldsymbol{a}
                    \,+\,\tilde{\omega}_{q}\boldsymbol{b}^{\dagger}\boldsymbol{b}
                    \,-\,\frac{\chi_{qq}}{2}\boldsymbol{b}^{\dagger}\boldsymbol{b}^{\dagger}\boldsymbol{b}\boldsymbol{b}
                    \,-\,\chi_{qr}\boldsymbol{a}^{\dagger}\boldsymbol{a}\boldsymbol{b}^{\dagger}\boldsymbol{b}
                    \nonumber\\[1mm]
                &\ -\,\frac{\chi_{rr}}{2}\boldsymbol{a}^{\dagger}\boldsymbol{a}^{\dagger}\boldsymbol{a}\boldsymbol{a}
                \nonumber\\[1mm]
                &=\,\omega_{r}\boldsymbol{a}^{\dagger}\boldsymbol{a}
                    \,+\,\omega_{q}\boldsymbol{b}^{\dagger}\boldsymbol{b}
                    \,-\,\frac{\chi_{qq}}{2}\bigl(\boldsymbol{b}^{\dagger}\boldsymbol{b}\bigr)^{2}
                    \,-\,\chi_{qr}\boldsymbol{a}^{\dagger}\boldsymbol{a}\boldsymbol{b}^{\dagger}\boldsymbol{b}
                    \nonumber\\[1mm]
                &\ -\,\frac{\chi_{rr}}{2}\bigl(\boldsymbol{a}^{\dagger}\boldsymbol{a}\bigr)^{2}\,,
\end{align}
where $\omega_{r}=\tilde{\omega}_{r}+\chi_{rr}/2$ and $\omega_{q}=\tilde{\omega}_{q}+\chi_{qq}/2$. The dressed qubit anharmonicity $\chi_{qq}$, the qubit-cavity cross-Kerr coefficient $\chi_{qr}$, and the cavity self-Kerr coefficient $\chi_{rr}$ can be expressed directly via the coupling strength~$g$ and the detuning~$\Delta$ as
\begin{subequations}
\begin{align}
 \chi_{qq}  &=\,\frac{\beta}{4}\left(1\,+\,\frac{|\Delta|}{\sqrt{\Delta^{2}+4g^{2}}}\right)^{2}\,,\\[1mm]
 \chi_{qr}  &=\,\frac{\beta}{2}\frac{4g^{2}}{\Delta^{2}+4g^{2}}\,,\\[1mm]
 \chi_{rr}  &=\,\frac{\beta}{4}\left(1\,-\,\frac{|\Delta|}{\sqrt{\Delta^{2}+4g^{2}}}\right)^{2}\,.
\end{align}
\end{subequations}

\hypertarget{sec:two transmons coupled to a cavity2}{}
\subsection{Two transmons coupled to a cavity}\label{sec:two transmons coupled to a cavity}
\bookmark[named=sec:two transmons coupled to a cavity,level=0,dest=sec:two transmons coupled to a cavity2]{Appendix A. III. Two transmons coupled to a cavity}

When two transmon qubits are coupled to the cavity, we may write an analogue expression to Eq.~(\ref{eq:single transmon coupled to resonator rot wave approx}). That is, in the rotating wave approximation we have the Hamiltonian
\begin{align}
  H_{qqr}/\hbar &=\, \omega_{R}\,\boldsymbol{c}^{\dagger}\boldsymbol{c}
    \,+\,\omega_{Q_{1}}\,\boldsymbol{q}_{1}^{\dagger}\boldsymbol{q}_{1}
    \,+\,\omega_{Q_{2}}\,\boldsymbol{q}_{2}^{\dagger}\boldsymbol{q}_{2}
    \nonumber\\[1mm]
    &\  +\,g_{1}\left(\boldsymbol{c}^{\dagger}\boldsymbol{q}_{1}\,+\,\boldsymbol{c}\hspace*{0.5pt}\boldsymbol{q}_{1}^{\dagger}\right)
        \,+\,g_{2}\left(\boldsymbol{c}^{\dagger}\boldsymbol{q}_{2}\,+\,\boldsymbol{c}\hspace*{0.5pt}\boldsymbol{q}_{2}^{\dagger}\right)
    \nonumber\\[1mm]
    &\ -\,\frac{\beta_{1}}{2}\boldsymbol{q}_{1}^{\dagger}\boldsymbol{q}_{1}^{\dagger}\boldsymbol{q}_{1}\boldsymbol{q}_{1}
        \,-\,\frac{\beta_{2}}{2}\boldsymbol{q}_{2}^{\dagger}\boldsymbol{q}_{2}^{\dagger}\boldsymbol{q}_{2}\boldsymbol{q}_{2}\,,
\end{align}
where we have neglected any direct coupling of the qubits to each other. In the two-qubit case (and beyond), an analytical diagonalization of the harmonic part (terms quadratic in the mode operators) becomes infeasible. However, using numerical methods and following similar arguments as presented in the previous Sec.~\hyperref[sec:Transmons Coupled to a Resonator]{A.II} one arrives at the effective Hamiltonian
\begin{align}
    H_{qqr}/\hbar   &=\,\omega_{r}\boldsymbol{a}^{\dagger}\boldsymbol{a}\,+\,\sum\limits_{i}\omega_{q_{i}}\bdn{i}\bn{i}\,
    -\,\sum\limits_{i}\chi_{q_{i}r}\,\boldsymbol{a}^{\dagger}\boldsymbol{a}\,\bdn{i}\bn{i}
    \,\nonumber\\
    &\ \ -\,\sum\limits_{i}\frac{\chi_{q_{i}q_{i}}}{2}\bigl(\bdn{i}\bn{i}\bigr)^{\!2}\,
    -\,\frac{\chi_{rr}}{2}\bigl(\boldsymbol{a}^{\dagger}\boldsymbol{a}\bigr)^{\!2}\,.
    \label{eq:supercond qubits Hamiltonian supplemental}
\end{align}
Assuming that the transmons can be fabricated such that $\chi_{q_{i}q_{i}}\equiv\chi_{qq}$ for all qubits, we have numerically checked that the cavity self-Kerr coefficient $\chi_{rr}$ can be approximated as
\begin{align}
    \chi_{rr}   &\approx\frac{\chi_{q_{1}r}^{2}+\chi_{q_{2}r}^{2}}{4\chi_{qq}}\,.
    \label{eq:two qubit cavity anharmonicity approximation}
\end{align}


\hypertarget{sec:simulations2}{}
\section{Simulations of coherent controlization using transmon qubits}\label{sec:simulations}
\bookmark[named=sec:simulations,level=-1,dest=sec:simulations2]{Appendix B: Simulations of coherent controlization using transmon qubits}

In this section we present the simulations of our protocol for coherent controlization with two and three qubits. In the simulations, which were coded in PYTHON using the QuTiP library~\cite{JohanssonNationNori:QuTip2:2013}, we discuss the influence of the cavity Kerr effect, including the (partial) correction of the linear Kerr effect by way of reference frame adjustments, and the possibility for correcting it entirely using photon-number selective gates~\cite{HeeresVlastakisHollandKrastanovAlbertFrunzioJiangSchoelkopf2015}. In addition, all simulations assume the presence of amplitude and phase damping for the qubits, and photon loss in the resonator. The unconditional displacements can safely be assumed to be perfect. We further assume the single-qubit operations to be perfect, given that the pulses are slow enough to address only the zero-photon subspace, but short enough to fit within the $\Delta t$ time intervals of our protocol.

\hypertarget{sec:setup for simulations2}{}
\subsection{Setup for simulations}\label{sec:setup for simulations}
\bookmark[named=sec:setup for simulations,level=0,dest=sec:setup for simulations2]{Appendix B. I. Setup for simulations}

For the simulation we truncate the Hilbert space of the resonator to be spanned by the Fock states~$\ket{n}_{r}$ of photon numbers $n=0,1,\ldots,100$. This is~a good approximation since the maximal average photon number of the coherent states in our protocol is $\bar{n}_{max}=|2\alpha|^{2}=14$ ($\bar{n}_{max}=|4\alpha|^{2}=32$) photons for $2$ ($3$) qubits and the overlap with photon numbers larger than $100$ is hence negligible. For the Hilbert space of the transmon qubits we will each only consider the lowest two eigenstates, i.e., the qubit levels. The effective Hamiltonian in the frame rotating with $\omega_{r}$ and $\omega_{q}$ for the resonator and qubit Hilbert spaces, respectively, is given by
\begin{align}
    H_{\hspace*{-0.5pt}I,K}/\hbar   &=\,\bigl(H_{\hspace*{-0.5pt}I}+H_{\hspace*{-0.5pt}K}\bigr)/\hbar
    \nonumber\\[1mm]
    &=\,
    -\,\sum\limits_{i}\chi_{q_{i}r}\,\boldsymbol{a}^{\dagger}\boldsymbol{a}\,\bdn{i}\bn{i}\,
    -\,\frac{\chi_{rr}}{2}\bigl(\boldsymbol{a}^{\dagger}\boldsymbol{a}\bigr)^{\!2}
    \nonumber\\[1mm]
    &=\,
    -\,\sum\limits_{i}\chi_{q_{i}r}\,\boldsymbol{a}^{\dagger}\boldsymbol{a}\,\sigma^{+}_{i}\sigma^{-}_{i}\,
    -\,\frac{\chi_{rr}}{2}\bigl(\boldsymbol{a}^{\dagger}\boldsymbol{a}\bigr)^{\!2}\,,
    \label{eq:supercond qubits effective Hamiltonian}
\end{align}
where $\sigma^{\pm}_{j}=\sigma_{j}^{x}\pm i\sigma_{i}^{y}$ are the raising/lowering operators of the $i$-th qubit. In addition to the free time evolution we will include corrections for the linear order of the Kerr effect by including rotations $U_{\varphi}(\gamma)=\exp(-i\varphi_{\gamma}\boldsymbol{a}^{\hspace*{-0.5pt}\dagger}\!\boldsymbol{a})$ after each period of time evolution, where $\gamma$ is the maximal displacement of the different coherent state components in the preceding step of the protocol. The conditional single-qubit operations will be represented by
\begin{align}
    \ket{0}\!\bra{0}_{r}\otimes U_{q}(\theta_{i})\,+\,\Bigl(\mathds{1}_{r}-\ket{0}\!\bra{0}_{r}\Bigr)\otimes\mathds{1}_{q}\,,
    \label{eq:single qubit operations}
\end{align}
where the single-qubit $Y$-rotations on the second (and third) of the two (three) qubits are realized by~a time-dependent drive $U(\theta_{j})=\exp(-i\tfrac{\theta_{j}}{2}\sigma^{y}\tfrac{t}{T})$. The durations~$T$ of these drives are taken to be the durations of the waiting periods in the corresponding protocol steps ($T=\Delta t$ for two qubits). On top of the unitary time evolution, unconditional displacements, and conditional qubit operations, all of which will be assumed to be perfect, we will consider decoherence in the system. In particular, we assume that the dynamics of the overall state $\rho$ of the joint cavity-qubit system during the waiting periods of the protocol is governed by the master equation
\vspace*{-2mm}
\begin{align}
 \frac{d\rho}{dt} &=\,-\frac{i}{\hbar}\left[H_{\hspace*{-0.5pt}I,K}, \rho\right]
    \,+\,\frac{1}{2\tau_{r}}\left(2\boldsymbol{a}\rho\boldsymbol{a}^{\dagger}
    \,-\,\boldsymbol{a}^{\dagger}\boldsymbol{a}\rho
    \,-\,\rho\boldsymbol{a}^{\dagger}\boldsymbol{a}\right)
    \nonumber\\
   & + \frac{1}{2\tau_{q}}\sum\limits_{i}\left(2\boldsymbol{\sigma}_{i}^{-}\rho\hspace*{1pt} \boldsymbol{\sigma}_{i}^{+} - \boldsymbol{\sigma}_{i}^{+} \boldsymbol{\sigma}_{i}^{-}\rho - \rho\hspace*{1pt} \boldsymbol{\sigma}_{i}^{+} \boldsymbol{\sigma}_{i}^{-}\right)\nonumber\\
  & + \Bigl(\frac{1}{2\tau_{\phi}}-\frac{1}{4\tau_{q}}\Bigr)\sum\limits_{i}\left(\boldsymbol{\sigma}_{i}^{z}\rho\hspace*{1pt} \boldsymbol{\sigma}_{i}^{z} - \rho \right)\,.
\end{align}
As the output of interest of the simulations we consider the fidelities $\mathcal{F}_{r}$ for the resonator, i.e., the squared overlap with the vacuum at the final step of the protocol, given by $\mathcal{F}_{r}=_{\ r\!\!}\!\bra{0}\bigl(\tr_{q}\rho\bigr)\ket{0}_{r}$, and~$\mathcal{F}_{q}$ for the qubits, i.e., the squared overlap with the target state $\ket{\psi}_{q}$, given by $\mathcal{F}_{q}=_{\ q\!\!}\!\bra{\psi}\bigl(\tr_{r}\rho\bigr)\ket{\psi}_{\hspace*{-1pt}q}$. We evaluate these two fidelities separately because the success of the protocol is ultimately only determined by achieving a large overlap of the reduced qubits state with the desired target state, irrespective of the final cavity state. On the other hand, the cavity fidelity gives a more detailed overview of the errors incurred by the Kerr effect in the resonator. Therefore, both $\mathcal{F}_{r}$ and $\mathcal{F}_{q}$ are quantities of interest for our protocol. In addition to the final fidelities, we include plots of the Wigner function $W(x,y)$, where
\vspace*{-2mm}
\begin{align}
    W(x,y)  &=\,\frac{1}{2\pi\hbar}\int\limits_{-\infty}^{\infty}dy\,e^{ipy/\hbar}\,_{r\!\!}\bra{x-\tfrac{y}{2}}\bigl(\tr_{q}\rho\bigr)\ket{x+\tfrac{y}{2}}_{r}\,,
\nonumber
\end{align}
of the reduced resonator state $\rho_{r}=\tr_{q}\rho$, where $\ket{x}$ are the eigenstates of the quadrature operator~$\boldsymbol{\Phi}$ with eigenvalue~$x$, and plots of the absolute values of the density matrix elements $|_{q\!\!}\bra{\!\mu\nu\!}\rho_{q}\ket{\!mn\!}_{\!q}\!|$ of the reduced qubit state $\rho_{q}=\tr_{r}\rho$ throughout the steps of our protocol.

\hypertarget{sec:simulations for two qubits2}{}
\subsection{Simulations for two qubits}\label{sec:simulations for two qubits}
\bookmark[named=sec:simulations for two qubits,level=0,dest=sec:simulations for two qubits2]{Appendix B. II. Simulations for two qubits}

\begin{figure*}[ht!]
\hspace*{-12mm}
\begin{minipage}{0.46\textwidth}
    \includegraphics[width=1\textwidth]{2QB_protocol_schematic_ideal.pdf}\hspace*{2mm}
\end{minipage}
\begin{minipage}{0.47\textwidth}
\vspace*{-1mm}
\begin{tabular}{| c || c | c | c | c | c | c | c | c || c | c |}
\hline
  \# &  $\chi_{qr}$ [MHz] & $\alpha$ & $\tau_{q}$~[$\mu$s] & $\tau_{\phi}$~[$\mu$s] & $\ket{\xi}_{q} $ & $\theta_{1}$ & $\theta_{2}$ & $\theta_{3}$ & $\mathcal{F}_{r}$ & $\mathcal{F}_{q} $\\
  \hline\hline
  $(\textbf{c})$ & $3\times2\pi$ & $\sqrt{2}$ & 20 & 30 & $\ket{00}$ & $\tfrac{\pi}{2}$ & $\pi$ & $\pi$ & 0.907 & 0.916\\[0.7mm]
  \hline
  $(\textbf{d})$ & $1.5\times2\pi$ & $\sqrt{3}$ & 100 & 100 & $\ket{00}$ & $\tfrac{\pi}{2}$ & $\tfrac{\pi}{6}$ & $\tfrac{\pi}{3}$ & 0.938 & 0.953\\[0.7mm]
  \hline
  $(\textbf{e})$ & $1.5\times2\pi$ & $\sqrt{3}$ & 100 & 100 & $\ket{00}$ & $0$ & $0$ & $0$ & 0.943 & 1.000\\[0.7mm]
  \hline
  $(\textbf{f})$ & $1.5\times2\pi$ & $\sqrt{3}$ & 100 & 100 & $\ket{00}$ & $\tfrac{\pi}{12}$ & $\tfrac{\pi}{4}$ & $\pi$ & 0.955 & 0.953\\[0.7mm]
  \hline
  $(\textbf{g})$ & $1.5\times2\pi$ & $\sqrt{3}$ & 100 & 100 & $\ket{10}$ & $\tfrac{\pi}{4}$ & $\pi$ & $\tfrac{\pi}{6}$ & 0.943 & 0.959\\[0.7mm]
  \hline
\end{tabular}
\end{minipage}
\vspace*{-3mm}
\flushleft{\textbf{(a)}}\hspace*{0.44\textwidth}\textbf{(b)}\\
\vspace*{-1.5mm}\hspace*{1mm}
\begin{minipage}{0.49\textwidth}
    \includegraphics[width=1\textwidth,trim={6cm 0 7cm 0},clip]{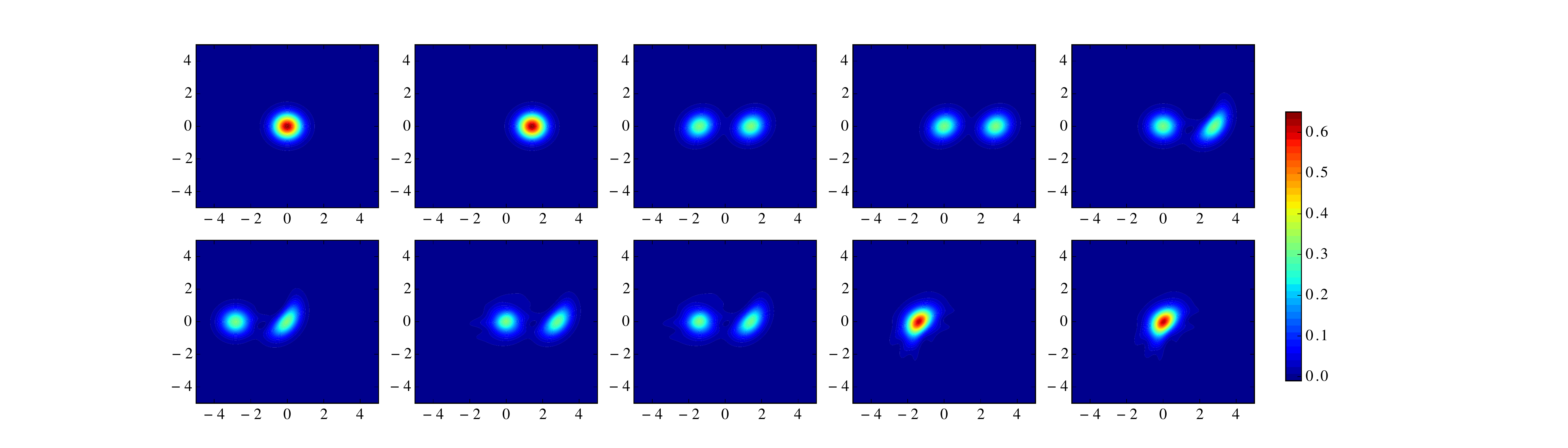}
\end{minipage}
\begin{minipage}{0.46\textwidth}
    \includegraphics[width=1\textwidth,trim={5cm 0 7cm 0},clip]{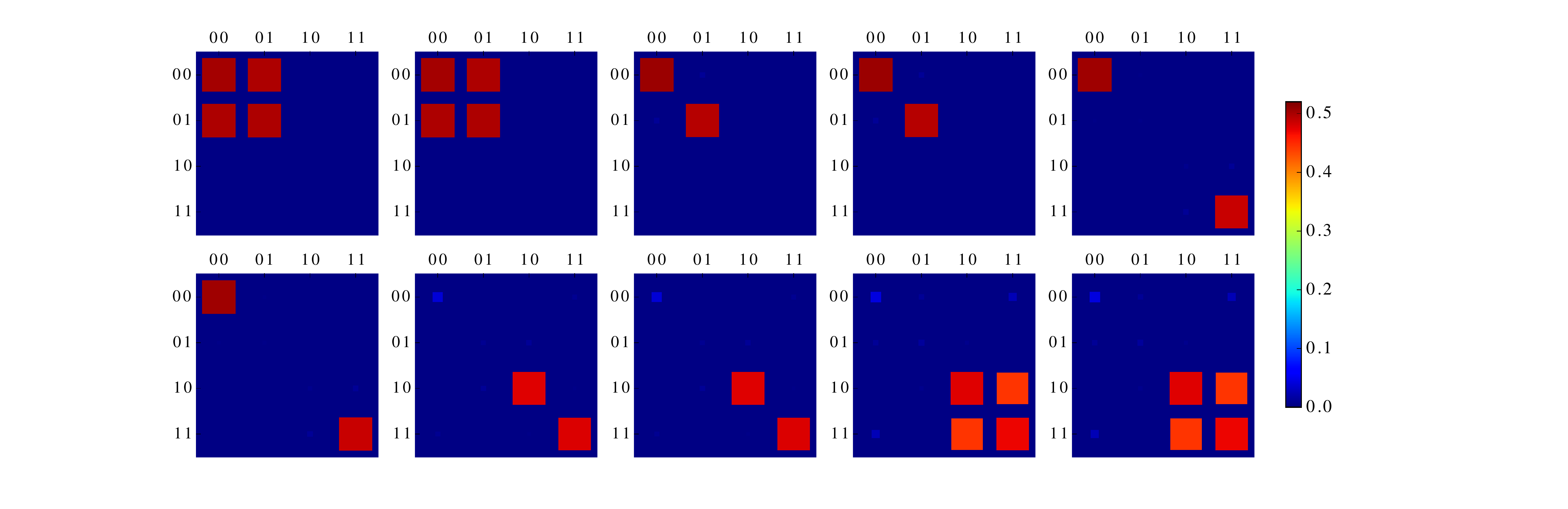}
\end{minipage}
\vspace*{-5mm}
\flushleft{\textbf{(c.1)}}\hspace*{0.47\textwidth}\textbf{(c.2)}
\vspace*{-2mm}\hspace*{1mm}
\begin{minipage}{0.49\textwidth}
    \includegraphics[width=1\textwidth,trim={6cm 0 7cm 0},clip]{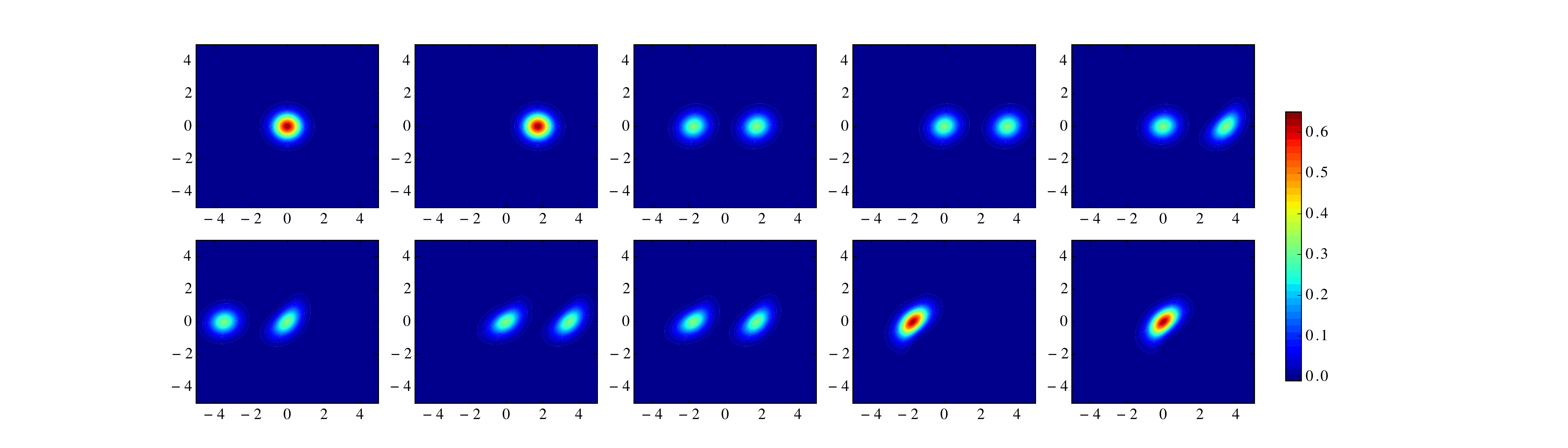}
\end{minipage}
\begin{minipage}{0.46\textwidth}
    \includegraphics[width=1\textwidth,trim={5cm 0 7cm 0},clip]{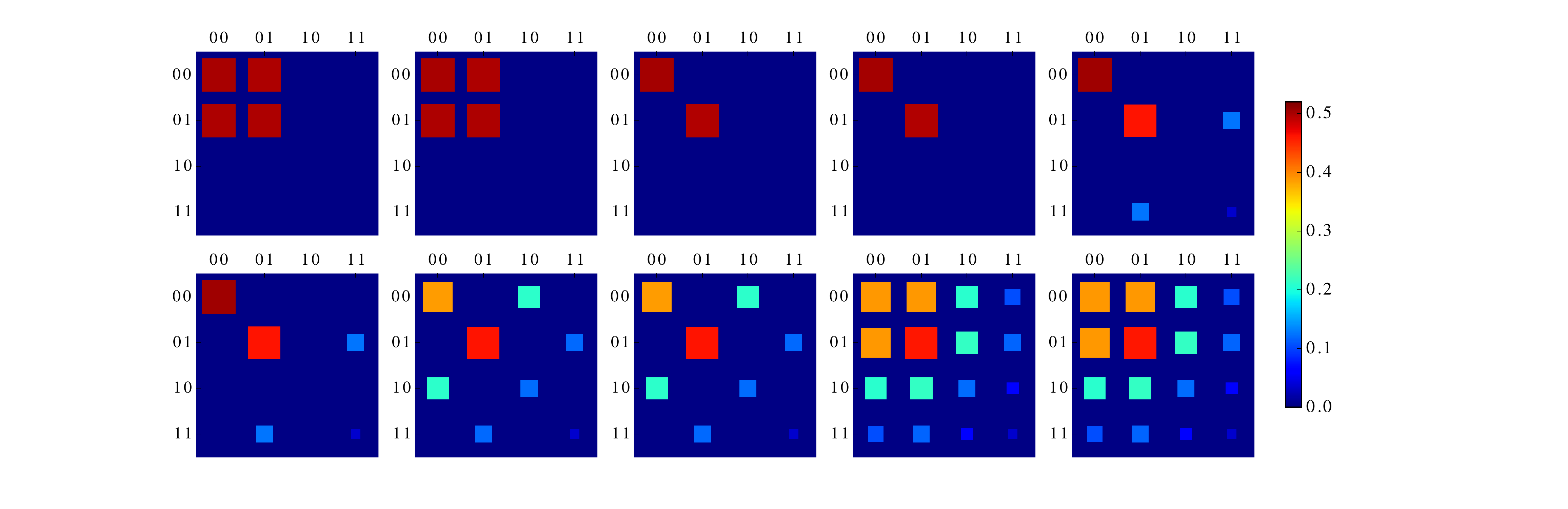}
\end{minipage}
\vspace*{-4mm}
\flushleft{\textbf{(d.1)}}\hspace*{0.47\textwidth}\textbf{(d.2)}
\vspace*{-2mm}\hspace*{1mm}
\begin{minipage}{0.49\textwidth}
    \includegraphics[width=1\textwidth,trim={6cm 0 7cm 0},clip]{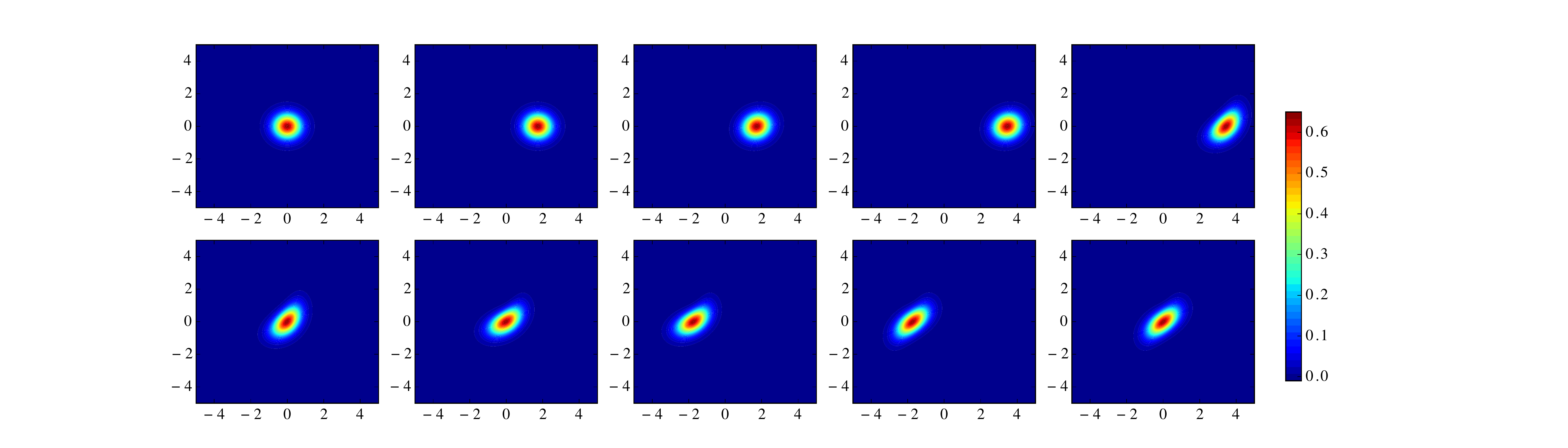}
\end{minipage}
\begin{minipage}{0.46\textwidth}
    \includegraphics[width=1\textwidth,trim={5cm 0 7cm 0},clip]{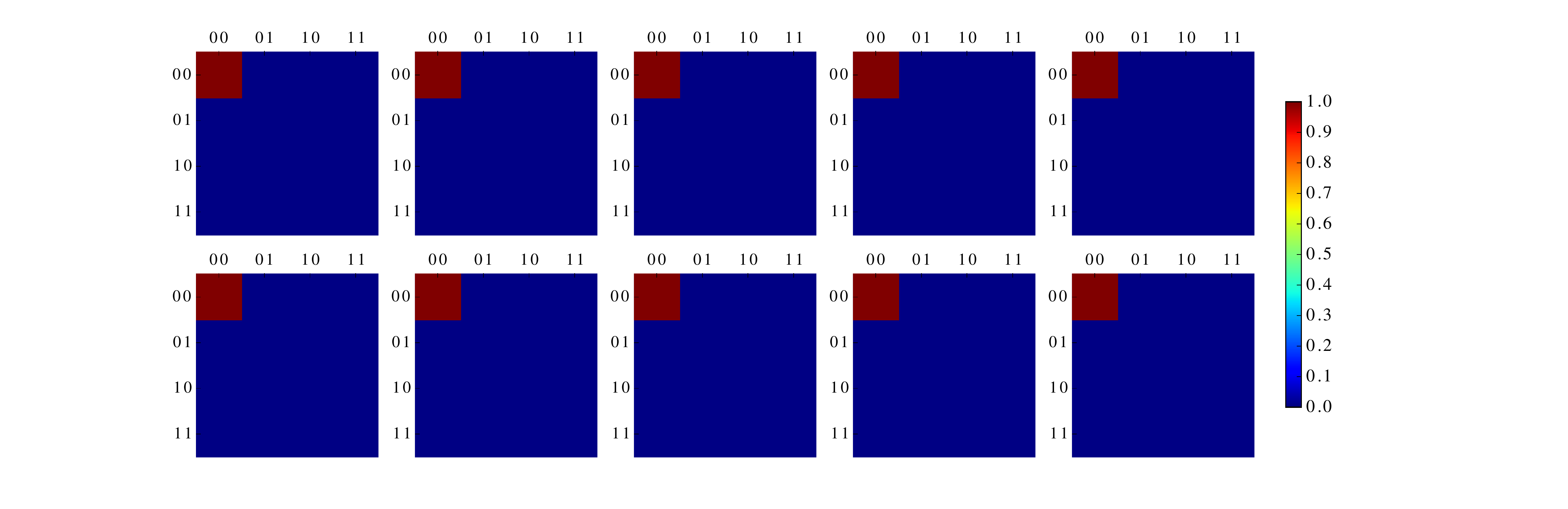}
\end{minipage}
\vspace*{-4mm}
\flushleft{\textbf{(e.1)}}\hspace*{0.47\textwidth}\textbf{(e.2)}
\vspace*{-2mm}\hspace*{1mm}
\begin{minipage}{0.49\textwidth}
    \includegraphics[width=1\textwidth,trim={6cm 0 7cm 0},clip]{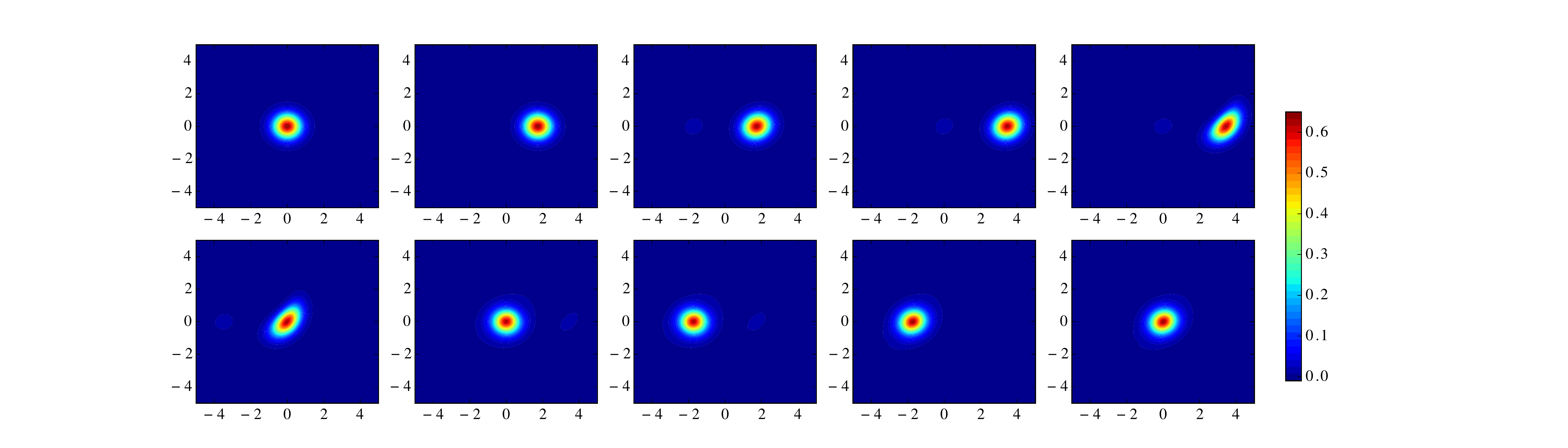}
\end{minipage}
\begin{minipage}{0.46\textwidth}
    \includegraphics[width=1\textwidth,trim={5cm 0 7cm 0},clip]{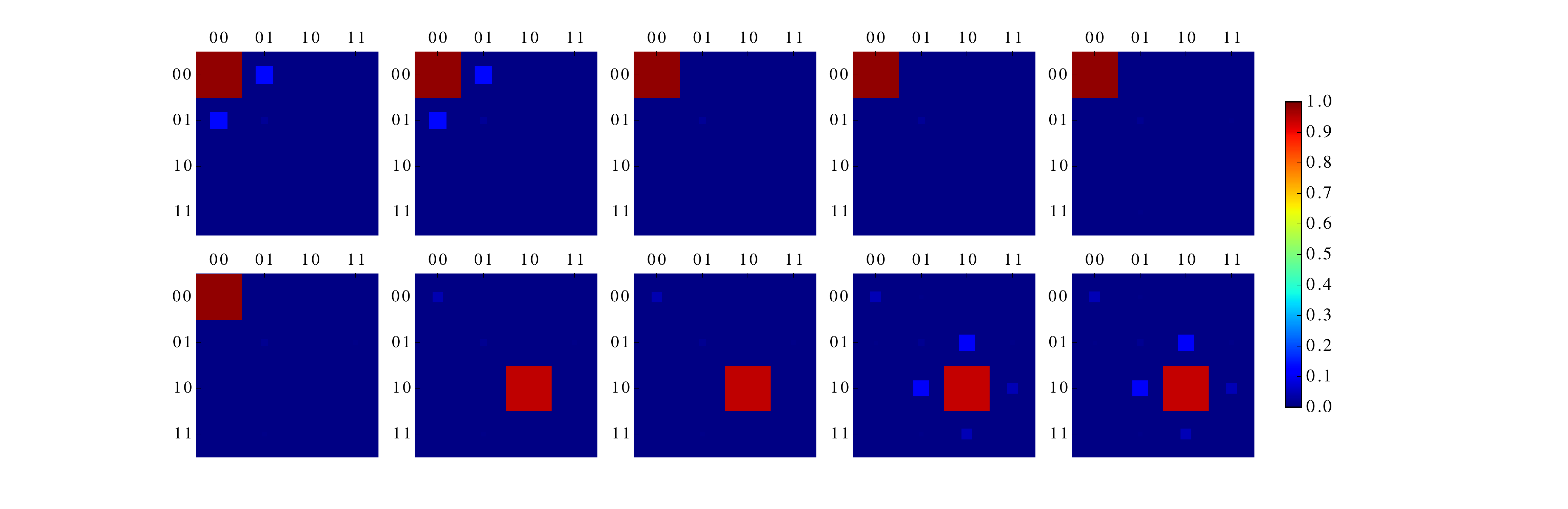}
\end{minipage}
\vspace*{-4mm}
\flushleft{\textbf{(f.1)}}\hspace*{0.47\textwidth}\textbf{(f.2)}
\vspace*{-2mm}\hspace*{1mm}
\begin{minipage}{0.49\textwidth}
    \includegraphics[width=1\textwidth,trim={6cm 0 7cm 0},clip]{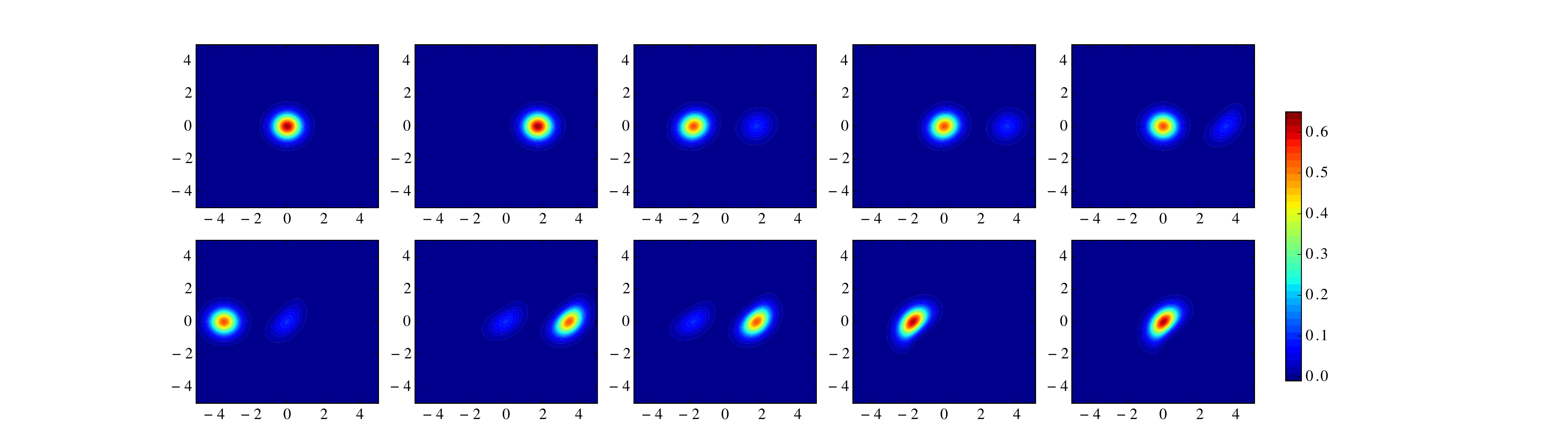}
\end{minipage}
\begin{minipage}{0.46\textwidth}
    \includegraphics[width=1\textwidth,trim={5cm 0 7cm 0},clip]{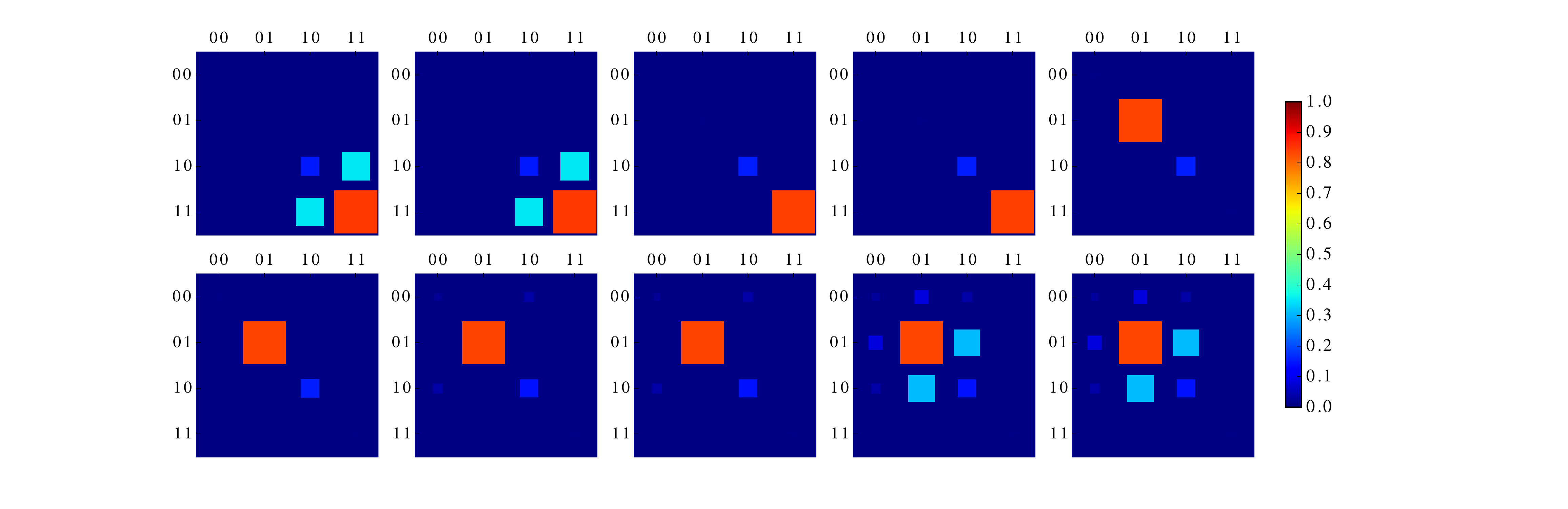}
\end{minipage}
\vspace*{-4mm}
\flushleft{\textbf{(g.1)}}\hspace*{0.47\textwidth}\textbf{(g.2)}
    \caption{\label{fig:two qubit simulation results}
    \textbf{Two-qubit simulations}: \textbf{(a)} Ideal protocol. \textbf{(b)} Table of simulation results
    . \textbf{(c)}-\textbf{(g)} show the reduced resonator state Wigner function on the left-hand side, and the absolute values of the density matrix elements $|_{q\!\!}\bra{\!\mu\nu\!}\rho_{q}\ket{\!mn\!}_{\!q}\!|$ of the reduced qubit states on the right-hand side. }
\end{figure*}

For two qubits, the qubit-cavity cross-Kerr coefficients $\chi_{qr}\equiv\chi_{q_{2}r}=\chi_{q_{1}r}/2$ are chosen in~a range from $1.5\times2\pi$~MHz to $3\times2\pi$~MHz. According to the approximation in Eq.~(\ref{eq:two qubit cavity anharmonicity approximation}), we further set $\chi_{rr}=5\chi_{qr}^{2}/(4\chi_{qq})$, where $\chi_{qq}$ is fixed to $300\times2\pi~$MHz. The coherence time for the resonator is set to $\tau_{r}=100~\mu$s throughout. For the simulation results presented in Fig.~\ref{fig:two qubit simulation results}, we hence vary the qubit-cavity coupling, the displacements in the protocol, the decoherence times, the initial qubit state $\rho_{q}=\ket{\xi}\!\bra{\xi}_{q}$ and the angles $\theta_{j}$ $(j=1,2,3)$ for the single-qubit operations. In addition to the results shown in Fig.~\ref{fig:two qubit simulation results}, we have simulated the influence of decoherence on the protocol when assuming that the cavity Kerr effect has been fully corrected using the methods of~\cite{HeeresVlastakisHollandKrastanovAlbertFrunzioJiangSchoelkopf2015}. In that case, using the same parameters as for the simulation
shown in Fig.~\ref{fig:coherent controlization Kerr corrected}~(c,d) of the main text, but setting $\chi_{rr}\equiv0$, we obtain $\mathcal{F}_{r}=99\%$ and $\mathcal{F}_{q}=96\%$.

Finally, note that the parameters $\chi_{q_{i}r}$, $\chi_{q_{i}q_{i}}$ and $\chi_{rr}$ are derived from the quantities $E_{C_{i}}$, $E_{J_{i}}$, $\omega_{R}$, and the cavity-qubit couplings $g_{i}$ for $i=1,2$. Achieving the exact required ratio $\chi_{q_{1}r}=2\chi_{q_{2}r}$ hence relies on the fine tuning of these quantities. For instance, to approximately achieve the parameters of the simulation in Fig.~\ref{fig:two qubit simulation results}~(c), one may set $E_{J_{1}}=27.0$~GHz, $E_{J_{2}}=20.1$~GHz, $E_{C_{1}}=E_{C_{2}}=0.3$~GHz, $g_{1}/(2\pi)=101$~MHz, $g_{2}/(2\pi)=127$~MHz, and $\omega_{R}/(2\pi)=9.16$~GHz to obtain $\chi_{q_{2}r}/(2\pi)=1.499$~MHz, $\chi_{q_{1}r}/(2\pi)=2.982$~MHz~$=1.989~\chi_{q_{2}r}$, $\chi_{q_{1}q_{1}}/(2\pi)=296.998$~MHz, $\chi_{q_{2}q_{2}}/(2\pi)=298.46$~MHz, and $\chi_{rr}/(2\pi)=9.367$~kHz (as compared to $\sum_{i}\chi_{q_{i}r}^{2}/(4\chi_{q_{i}q_{i}})=9.34$~kHz). For the simulations, we assume that such deviations are corrected by an echo-type operation as explained in Sec.~\ref{sec:other corrections} of the main text. Finally, note that all parameters used in the simulations are compatible with the dispersive approximation.

\hypertarget{sec:simulations for three qubits2}{}
\subsection{Simulations for three qubits}\label{sec:simulations for three qubits}
\bookmark[named=sec:simulations for three qubits,level=0,dest=sec:simulations for three qubits2]{Appendix B. III. Simulations for three qubits}

\begin{figure*}
\hspace{-14mm}\textbf{(a)}\includegraphics[width=0.875\textwidth]{3QB_protocol_schematic_ideal.pdf}\\
\hspace{5mm}\textbf{(b)}\includegraphics[width=0.98\textwidth,trim={8.2cm 1.5cm 10cm 2cm},clip]{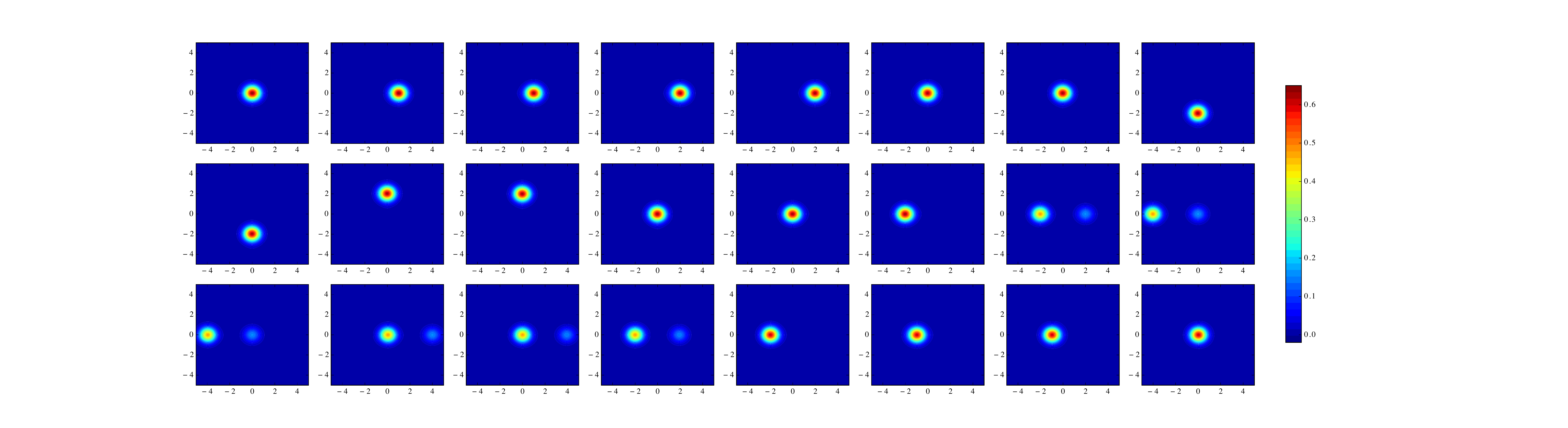}\\
\hspace{5mm}\textbf{(c)}\includegraphics[width=0.95\textwidth,trim={8.1cm 2.4cm 10cm 1.5cm},clip]{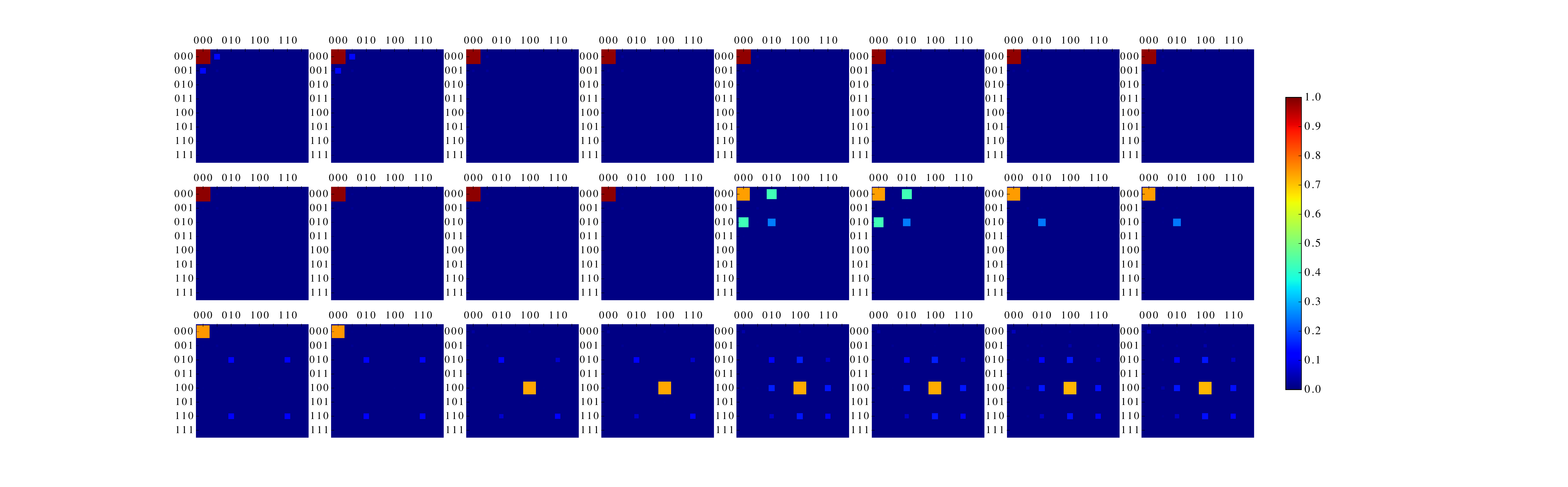}
\caption{\label{fig:three qubit simulation results s12}
    \textbf{Three-qubit simulation $s_{12}$}: \textbf{(a)} Ideal protocol. \textbf{(b)} The Wigner function of the reduced resonator state of the three-qubit protocol is shown for the parameters of the simulation specified in $s_{12}$ of Table~\ref{table:three qubit simulation results}, where the cavity Kerr effect is assumed to be fully corrected. \textbf{(c)} shows the corresponding plots of the absolute values of the density matrix elements $|_{q\!\!}\bra{\!\mu\nu\lambda\!}\rho_{q}\ket{\!mnl\!}_{\!q}\!|$ ($\mu,\nu,\lambda,m,n,l=0,1$) of the reduced qubit states.}
\end{figure*}

For three qubits, the qubit-cavity cross-Kerr coefficients $\chi_{qr}\equiv\chi_{q_{3}r}=2\chi_{q_{2}r}=4\chi_{q_{1}r}$ are chosen in~a range from $0.2\times2\pi$~MHz to $0.4\times2\pi$~MHz. According to the approximation in Eq.~(\ref{eq:two qubit cavity anharmonicity approximation}), we further set $\chi_{rr}=21\chi_{qr}^{2}/(4\chi_{qq})$, where $\chi_{qq}$ is fixed to $300\times2\pi~$MHz. The coherence times for the resonator and the qubit (dephasing and amplitude damping) are set to $100~\mu$s. For the simulation results presented in Table~\ref{table:three qubit simulation results}, we hence vary the qubit-cavity coupling, the displacements in the protocol, the initial qubit state $\rho_{q}=\ket{\xi}\!\bra{\xi}_{q}$ and the angles $\theta_{j}$ $(j=1,2,\ldots,7)$ for the single-qubit operations. A selection of these simulations Figs.~\ref{fig:three qubit simulation results s11} and~\ref{fig:three qubit simulation results s12} further illustrate the reduced states of the resonator and the qubits throughout the protocol.

\begin{table}[hb!]
\begin{tabular}{| c || c | c | c | c | c | c | c | c | c | c || c | c |}
  \hline
  \# &  $\chi_{qr}$~[MHz] & $\alpha$ & $\ket{\xi}_{q} $ & $\theta_{1}$ & $\theta_{2}$ & $\theta_{3}$ &
  $\theta_{4}$ & $\theta_{5}$ & $\theta_{6}$ & $\theta_{7}$ & $\mathcal{F}_{r}$ & $\mathcal{F}_{q}$ \\
  \hline
  $s_{1}$ & $0.3\times2\pi$ & $1$ & $\ket{\xi_{u}}$ & 0 & 0 & 0 & 0 & 0 & 0 & 0 & 0.671 & 0.296 \\[0.7mm]
  \hline
  $s_{2}$ & $0.3\times2\pi$ & $1$ & $\ket{\!010\!}$ & $\tfrac{\pi}{2}$ & $\pi$ & 0 & 0 & $\pi$ & 0 & $\tfrac{\pi}{2}$ & 0.505 & 0.449 \\[0.7mm]
  \hline
  $s_{3}$ & $0.4\times2\pi$ & 1 & $\ket{\!000\!}$ & 0 & 0 & 0 & 0 & 0 & 0 & 0 & 0.832 & 1.000 \\[0.7mm]
    \hline
  $s_{4}$ & $0.3\times2\pi$ & 1 & $\ket{\!000\!}$ & $\tfrac{\pi}{2}$ & 0 & $\tfrac{\pi}{6}$ & 0 & $\pi$ & 0 & $\tfrac{\pi}{2}$ & 0.699 & 0.481 \\[0.7mm]
    \hline
  $s_{5}$ & $0.4\times2\pi$ & 1 & $\ket{\!000\!}$ & $\tfrac{\pi}{2}$ & 0 & $\tfrac{\pi}{6}$ & 0 & $\pi$ & 0 & $\tfrac{\pi}{2}$ & 0.578 & 0.457 \\[0.7mm]
    \hline
  $s_{6}$ & $0.2\times2\pi$ & 1 & $\ket{\!000\!}$ & $\tfrac{\pi}{2}$ & 0 & $\tfrac{\pi}{6}$ & 0 & $\pi$ & 0 & $\tfrac{\pi}{2}$ & 0.747 & 0.444 \\[0.7mm]
    \hline
  $s_{7}$ & $0.35\times2\pi$ & 1 & $\ket{\!000\!}$ & $\tfrac{\pi}{2}$ & 0 & $\tfrac{\pi}{6}$ & 0 & $\pi$ & 0 & $\tfrac{\pi}{2}$ & 0.646 & 0.475 \\[0.7mm]
    \hline
  $s_{8}$ & $0.3\times2\pi$ & 1 & $\ket{\!000\!}$ & $\pi$ & 0 & 0 & 0 & 0 & 0 & $\pi$ & 0.697 & 0.797 \\[0.7mm]
    \hline
  $s_{9}$ &  $0.3\times2\pi$ & $\sqrt{\tfrac{1}{2}}$ & $\ket{\!000\!}$ & $\pi$ & 0 & 0 & 0 & 0 & 0 & $\pi$ & 0.674 & 0.729 \\[0.7mm]
    \hline
  $s_{10}$ &  $0.3\times2\pi$ & $\sqrt{\tfrac{3}{2}}$ & $\ket{\!000\!}$ & $\pi$ & 0 & 0 & 0 & 0 & 0 & $\pi$ & 0.329 & 0.796 \\[0.7mm]
    \hline
  $s_{11}$ & $0.3\times2\pi$  & $1$ & $\ket{\!000\!}$ & $\tfrac{\pi}{12}$ & 0 & $\tfrac{\pi}{3}$ & 0 & 0 & $\tfrac{\pi}{2}$ & $\pi$ & 0.804 & 0.639 \\[0.7mm]
    \hline
  $s_{12}$ & $0.3\times2\pi$  & $1$ & $\ket{\!000\!}$ & $\tfrac{\pi}{12}$ & 0 & $\tfrac{\pi}{3}$ & 0 & 0 & $\tfrac{\pi}{2}$ & $\pi$ & 0.953 & 0.748 \\[0.7mm]
  \hline
\end{tabular}
\caption{\label{table:three qubit simulation results}
    \textbf{Three-qubit simulations}: The table shows the results of the numerical simulations for three qubits for various system parameters, rotation angles~$\theta_{j}$, and initial states, where $\ket{\xi_{u}}$ is the uniform superposition overall three qubit computational basis states. The last simulation, $s_{12}$ was executed for the same parameters as $s_{11}$, but assuming that the cavity Kerr effect has been corrected (here: $\chi_{rr}\equiv0$) using the photon-number selective phase gates~\cite{HeeresVlastakisHollandKrastanovAlbertFrunzioJiangSchoelkopf2015}.}
\end{table}

\newpage

\begin{figure*}
\hspace{-14mm}\textbf{(a)}\includegraphics[width=0.875\textwidth]{3QB_protocol_schematic_ideal.pdf}\\
\hspace{5mm}\textbf{(b)}\includegraphics[width=0.98\textwidth,trim={8.2cm 1.5cm 10cm 2cm},clip]{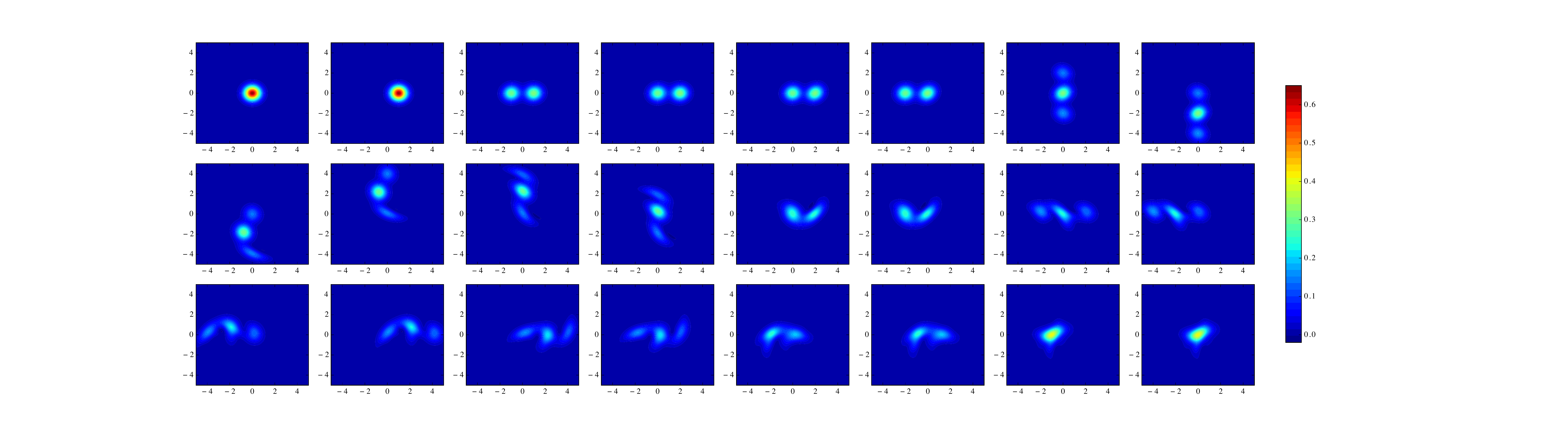}\\
\hspace{5mm}\textbf{(c)}\includegraphics[width=0.95\textwidth,trim={8.1cm 2.4cm 10cm 1.5cm},clip]{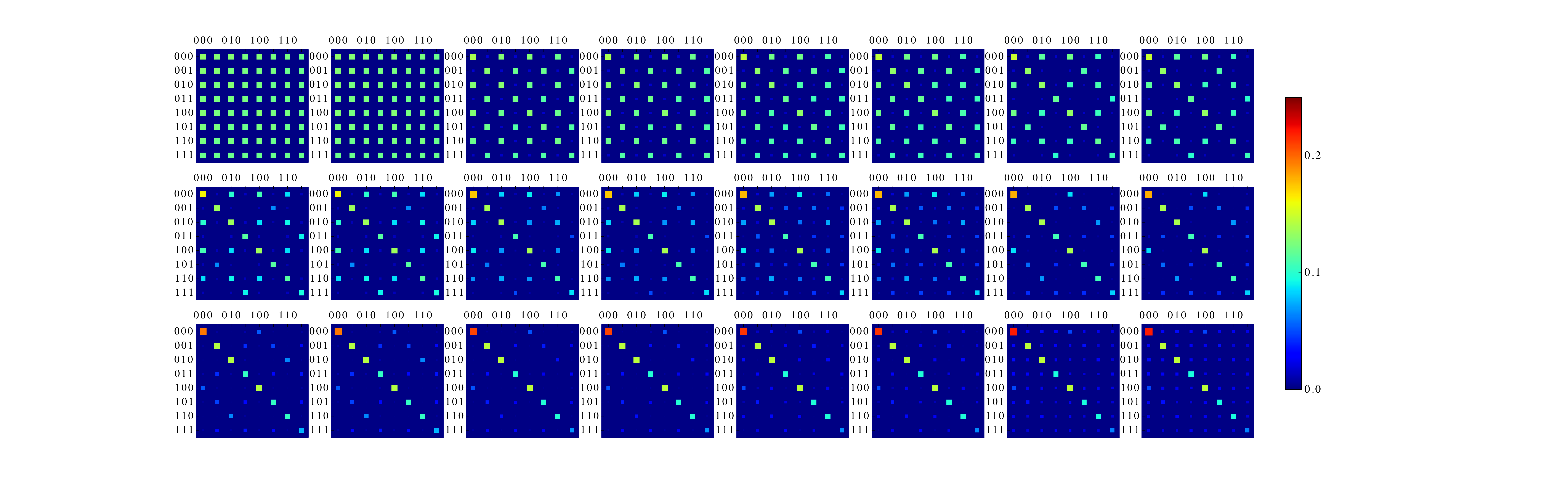}
\caption{\label{fig:three qubit simulation results s11}
    \textbf{Three-qubit simulation $s_{1}$}: \textbf{(a)} Ideal protocol. \textbf{(b)} The Wigner function of the reduced resonator state of the three-qubit protocol is shown for the parameters of the simulation specified in $s_{1}$ of Table~\ref{table:three qubit simulation results}. \textbf{(c)} shows the corresponding plots of the absolute values of the density matrix elements $|_{q\!\!}\bra{\!\mu\nu\lambda\!}\rho_{q}\ket{\!mnl\!}_{\!q}\!|$ ($\mu,\nu,\lambda,m,n,l=0,1$) of the reduced qubit states.}
\end{figure*}

\begin{figure*}
\hspace{-14mm}\textbf{(a)}\includegraphics[width=0.875\textwidth]{3QB_protocol_schematic_ideal.pdf}\\
\hspace{5mm}\textbf{(b)}\includegraphics[width=0.98\textwidth,trim={8.2cm 1.5cm 10cm 2cm},clip]{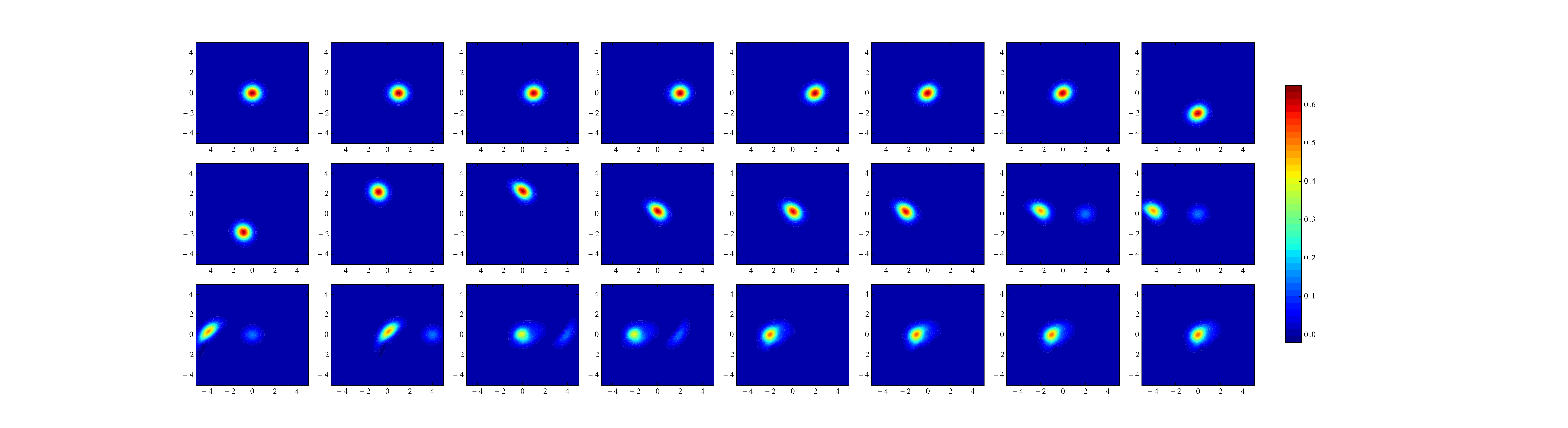}\\
\hspace{5mm}\textbf{(c)}\includegraphics[width=0.95\textwidth,trim={8.1cm 2.4cm 10cm 1.5cm},clip]{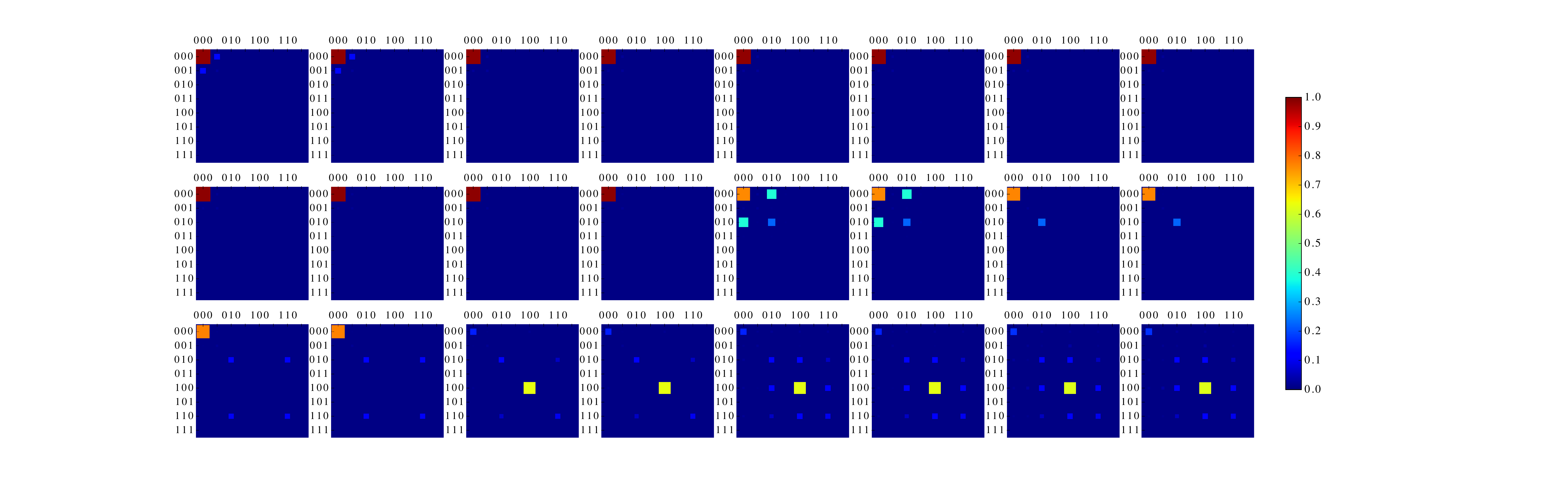}
\caption{\label{fig:three qubit simulation results s11}
    \textbf{Three-qubit simulation $s_{11}$}: \textbf{(a)} Ideal protocol. \textbf{(b)} The Wigner function of the reduced resonator state of the three-qubit protocol is shown for the parameters of the simulation specified in $s_{11}$ of Table~\ref{table:three qubit simulation results}. \textbf{(c)} shows the corresponding plots of the absolute values of the density matrix elements $|_{q\!\!}\bra{\!\mu\nu\lambda\!}\rho_{q}\ket{\!mnl\!}_{\!q}\!|$ ($\mu,\nu,\lambda,m,n,l=0,1$) of the reduced qubit states.}
\end{figure*}

\end{document}